\documentclass[3p,times]{elsarticle}

\usepackage{ecrc}

\volume{00}

\firstpage{1}

\journalname{Computer Networks}

\runauth{}

\jid{procs}

\jnltitlelogo{Computer Networks}

\CopyrightLine{2011}{Published by Elsevier Ltd.}

\usepackage{amssymb}

\usepackage[figuresright]{rotating}

\usepackage{amsmath,amsfonts,amssymb}
\usepackage{amsthm}
\usepackage{array}
\usepackage{stfloats}
\usepackage{verbatim}
\usepackage{graphicx}
\usepackage{url}
\usepackage{textcomp}
\usepackage{dirtytalk}
\usepackage{adjustbox}
\usepackage{comment}
\usepackage{ifoddpage}
\usepackage{tikz}
\usepackage{xcolor}
\usepackage{float}
\usepackage[font=normal,labelfont=normal]{caption}
\usepackage{svg}
\usepackage{pifont}
\usepackage{booktabs}
\usepackage{longtable}            
\usepackage{rotating}
\usepackage{multirow}
\usepackage{pifont}
\usepackage{listings} 
\usepackage{threeparttable}

\definecolor{keywordcolor}{rgb}{0,0,1}     
\definecolor{identifiercolor}{rgb}{0.5,0,0.5} 
\definecolor{commentcolor}{rgb}{0,0.5,0}   
\definecolor{stringcolor}{rgb}{0.8,0,0}    

\definecolor{keywordcolor}{rgb}{0,0,1} 
\definecolor{variablecolor}{rgb}{0.5,0,0.5} 
\definecolor{commentcolor}{rgb}{0,0.5,0} 

\lstdefinelanguage{Solidity}{
    keywords={contract, function, if, else, for, while, return, uint, uint256, address, mapping, require, string, public, view, memory, struct},
    keywordstyle=\color{keywordcolor}\bfseries,
    identifierstyle=\color{identifiercolor},
    comment=[l]{//},
    morecomment=[s]{/*}{*/},
    commentstyle=\color{commentcolor}\itshape,
    stringstyle=\color{stringcolor},
    sensitive=true
}

\usepackage{algorithm}
\usepackage{algpseudocode}
\usepackage{subfig}

\newcommand{\contract}[1]{\textit{#1}}    
\newcommand{\function}[1]{\texttt{#1}}

\usepackage{enumitem}

\usepackage{hyperref}

\begin{document}

\begin{frontmatter}


\title{Performance Analysis, Lessons Learned and Practical Advice for a 6G Inter-Provider DApp on the Ethereum Blockchain}


%

\dochead{}


\author{Farhana Javed and Josep Mangues-Bafalluy}

\address{Services as networkS (SaS) Research Unit, \\ Centre Tecnològic de Telecomunicacions de Catalunya (CTTC/CERCA), \\ Castelldefels, Spain \\ email: \{farhana.javed, josep.mangues\}@cttc.es}

\begin{abstract}
This paper presents a multi-contract blockchain framework for inter-provider agreements in 6G networks, emphasizing performance analysis under a realistic Proof-of-Stake (PoS) setting on Ethereum’s Sepolia testnet. We begin by quantifying Ethereum Virtual Machine (EVM)-based gas usage for critical operations such as provider registration, service addition, and SLA penalty enforcement, observing that cold writes and deep data structures can each inflate gas consumption by up to 20\%. We then examine block-level dynamics when multiple transactions execute concurrently, revealing that moderate concurrency (e.g., 30–50 simultaneous transactions) can fill blocks to 80–90\% of their gas limit and nearly double finalization times from around 15 seconds to over 30 seconds. Finally, we synthesize these insights into a practical design guide, demonstrating that flattening nested mappings, consolidating storage writes, and selectively timing high-impact transactions can markedly reduce costs and latency spikes. Collectively, our findings underscore the importance of EVM-specific optimizations and transaction scheduling for large-scale DApps in 6G telecom scenarios. The implementation of this work is publicly accessible \href{https://github.com/farhanajaved/6G-InterProvider-DApp-Performance-Lessons/tree/main}{online}.

\end{abstract}


\end{frontmatter}

\section{Introduction}
\label{sec1}

The next generation of mobile networks (6G) anticipates supporting diverse services by seamlessly scaling resources across multiple administrative domains. Network Function Virtualization (NFV) is foundational to this vision, as it virtualizes traditional network elements into modular software components. Through NFV, network resources can be managed flexibly and openly, enabling dynamic service orchestration, multi-vendor interoperability, and efficient sharing and use of infrastructure among independent stakeholders (e.g., initiatives such as CAMARA \cite{camara} and GSMA’s Open Gateway framework \cite{gsma2023ecosystem,GSMA2021OpenVirtualisedRAN,GSMA_APAC_OpenRAN}).

Yet, despite these advantages, this open architecture introduces coordination complexities and trust challenges among multiple stakeholders. When operators experience sudden demand surges—such as during large-scale events; they often rely on peer operators for rapid resource provisioning. To manage such dynamic interactions, \textit{inter-provider agreements} have become critical, enabling flexible resource sharing while maintaining costs and service quality \cite{ETSIGSPDL024,3GPPTS23.251}. However, due to the decentralized nature of these arrangements, traditional centralized trust mechanisms are insufficient, creating a need for alternative approaches to secure transactions and enforce contracts \cite{ETSI2020Applications}. For example, NGMN emphasizes incorporating trustworthiness by design in 6G—addressing transparency, privacy, reliability, resilience, and safety. It highlights decentralized systems such as blockchain for managing inter-provider agreements, demonstrating a clear focus on fostering trust in business interactions and multi-party collaboration scenarios \cite{NGMN2023}.

Blockchain and Distributed Ledger Technologies (DLTs) directly address these trust and coordination issues by offering a decentralized, immutable, and transparent transaction ledger. Within telecommunications, groups such as the TM Forum Catalyst Program and ETSI’s Permissioned Distributed Ledgers initiative have demonstrated blockchain’s ability to automate settlement processes, Quality of Service (QoS) monitoring, and Service Level Agreement (SLA) enforcement among multiple operators \cite{TMForumCatalystExample,ETSI2019PDLInterop,ETSI2021SmartContracts,ETSI2022SLA}. Moreover, CAMARA is leveraging blockchain to integrate telecom identifiers with decentralized Web3 services, enabling secure identity management and seamless transactions \cite{camara_blockchain}. These decentralized frameworks allow operators to transparently lease or share surplus network capacity, thereby reducing operational costs and fostering collaboration in an increasingly multi-vendor environment \cite{ETSIGSPDL024,3GPPTS23.251}.

Although previous studies, including our own, have established the feasibility of using blockchain to support inter-provider resource sharing and SLA enforcement \cite{javed2022blockchainPIMRC,javed2022blockchainglobecom,javed2024blockchainLetter,JavedNFVworkshop,javed2025empiricalsmartcontractslatency}, critical implementation challenges remain unresolved. Existing literature primarily provides conceptual frameworks or high-level architectural insights but does not offer systematic guidance on designing end-to-end Decentralized Applications (DApps) under realistic network conditions. Our recent empirical study \cite{javed2025empiricalsmartcontractslatency} contributed an initial statistical analysis of transaction latency, examining how variations in gas prices, block sizes, and transaction counts affect overall performance. However, that study focused on latency metrics and did not address deeper implementation challenges specific to Ethereum Virtual Machine (EVM)-based ecosystems—such as role-based interactions, function-level performance, and internal contract-design decisions—challenges that are crucial for comprehensive DApp deployments. This paper builds on and extends those findings by investigating these critical dimensions across various EVM-compatible environments, offering actionable design recommendations for DApp architectures in real-world blockchain networks.

Despite the advances in blockchain-based resource sharing and SLA enforcement, critical research questions remain unanswered in the existing literature. Prior work \cite{faisal2022design,antevski2023applying,TMForumCatalystExample,ETSI2022SLA}, including our own, has not systematically examined how to design DApps that support role-based interactions across multiple administrative domains, nor has it explored the impact of internal smart-contract design choices—such as cold versus warm storage, data structure selection, and indexing—on performance in EVM-compatible blockchain networks (Polygon, Avalanche, Binance Smart Chain, etc.). Notably, these networks share core EVM features (e.g., bytecode format and account structure), ensuring that design principles and techniques developed for Ethereum generally apply with minimal modification—particularly for Layer-2 solutions such as Polygon, which also rely on the EVM for smart contract execution. Furthermore, transaction ordering and the sequential execution of dependent operations remain under-studied. Additionally, while gas price volatility, mempool dynamics, block-size variability, and Proof-of-Stake (PoS) transaction inclusion policies are known to influence performance \cite{javed2025empiricalsmartcontractslatency}, their effects on different role-based smart contracts, particularly the highest gas-consuming functions (such as adding data on blockchain using arrays) and dependent smart contract transactions, have yet to be quantified. Finally, there is a lack of clear, actionable guidelines for optimizing EVM-level decisions, including storage management and structuring of complex data (arrays, mappings). Moreover, concurrency management—specifically, multiple interactions with smart contract functions—and network-level considerations for these complex smart contracts have not been explored.

Addressing these gaps requires an in-depth understanding of smart-contract behavior at the EVM layer—an understanding that extends beyond Ethereum to other EVM-compatible ecosystems. Therefore, this paper provides a comprehensive performance analysis of multiple smart contracts deployed on the EVM in the PoS Ethereum blockchain. By developing and empirically evaluating a modular, multi-contract DApp on the Sepolia Ethereum test network, we quantify the practical implications of EVM-level decisions and offer concrete guidelines for smart-contract implementation under realistic operational conditions.

Accordingly, the key contributions of this paper are as follows:
\begin{itemize}
    \item We implement a modular DApp architecture composed of six separate smart contracts (\contract{RegistrationAD.sol}, \contract{AddService.sol}, \contract{SelectService.sol}, \texttt{RegisterBreach.sol}, \contract{CalculatePenalty.sol}, \contract{TransferFunds.sol}). Each contract focuses on a distinct operational task, ensuring clear functional separation. 
    \item We systematically quantify and empirically analyze the function-level performance of these smart contracts under live blockchain network conditions, explicitly examining gas usage patterns influenced by internal contract design decisions—such as cold versus warm storage accesses, data-structure complexity, and storage array indexing—at the individual transaction level.
    \item  We further investigate performance challenges by examining key metrics—such as transaction latency and block-level gas utilization—under increasing concurrency. In addition, we explore how dependent transaction (see Table \ref{tab:blockchain_terms})  scenarios within a PoS consensus model influence overall system performance.
    \item Based on empirical findings, we propose concrete EVM-level design recommendations for strategic optimization. These guidelines address storage management, data structuring, and transaction scheduling, making them broadly applicable to EVM-based blockchains and PoS networks beyond Ethereum. They are particularly relevant for advanced 6G use cases such as inter-provider resource sharing.
\end{itemize}

The structure of this article is as follows: Section \ref{sec2} reviews existing literature on network management and blockchain networks and outlines the primary contributions of this study. Section \ref{sec3} provides background on inter-provider agreements and blockchain, explaining how blockchain can address the challenges introduced by such agreements. Section \ref{sec4} presents the architectural framework, proposed components and interfaces, and the design of the smart contracts, while also detailing the operational workflow of the proposed framework-based approach for inter-provider agreements. Section \ref{sec5} describes the experimental evaluation, presents the obtained results, and summarizes key findings, including lessons learned. Section \ref{lesson-learned} discusses practical suggestions and techniques based on the analysis presented in the previous sections. Finally, Section \ref{sec6} concludes the paper and discusses future directions.

\section{Related Work}
\label{sec2}
Blockchain technology has garnered significant attention for fostering trust and security across multiple industries, including telecommunications. Notably, the TM Forum has explored blockchain’s potential for automating contract management and revenue-sharing arrangements among operators \cite{tm_forum_blockchain}, while the CAMARA API initiative applies blockchain for secure and interoperable digital service ecosystems \cite{camara_blockchain}. In parallel, ETSI has led projects on blockchain-based mechanisms for digital identity verification \cite{etsi_pdl_017}, IoT network protection \cite{etsi_pdl_026}, and privacy and consent management \cite{etsi_pdl_023}. Together, these standardization efforts highlight how decentralized ledgers can streamline multi-party coordination in telecom contexts.

Beyond these initiatives, academic research on blockchain in 5G and evolving networks underscores improvements in security, privacy, and trust \cite{faisal2022beat, antevski2022federation, baskaran2023role, zeydan2022blockchain, xevgenis2025blockchain}. Various works demonstrate how blockchain can help enforce privacy compliance \cite{augusto2024sok, li2024security, hasan2024blockchain}, enable trustless resource-sharing and trading \cite{alshahrani2024enabling, tripi2024security}, and facilitate federated network slicing \cite{hafi2024split} or multi-domain integration \cite{han2024lightweight}. This literature showcases blockchain’s versatility in meeting stringent technical and governance needs for advanced wireless systems. However, much of it remains conceptual—focused on broad, high-level architectures or abstract security/privacy models rather than detailing how specific on-chain operations behave under real usage patterns.

A research thread also examines scalability and performance in telecom and cloud environments, exploring how blockchain might handle large transaction volumes or distributed resource orchestration. For instance, \cite{zahir2024performance} analyzes blockchain-based multi-cloud solutions, while \cite{antevski2023applying} benchmarks different consensus algorithms across heterogeneous settings. Additional studies propose adjusting block size to reduce latency or discuss the economic trade-offs of deploying blockchain as on-premises, IaaS, or BaaS \cite{wilhelmi2022end, afraz2023blockchain}. Further, emerging work on 6G-oriented solutions explores novel consensus techniques and resource-management schemes tailored to next-generation network demands \cite{suetor2025overview}.

Despite these wide-ranging investigations, a gap persists when it comes to granular, function-level analyses of smart contracts under realistic conditions. Many studies focus on overall metrics—throughput, consensus overhead, or block utilization—without dissecting individual contract operations or exploring how concurrency (e.g., simultaneous registrations, multiple dependent function calls) impacts both gas consumption and transaction latency on public networks like Ethereum’s Sepolia. For example, while \cite{zahir2024performance} and \cite{antevski2023applying} reveal key performance trade-offs, they do not address the role-based workflows inherent in telecom-specific DApps or highlight how concurrency can strain block space. Similarly, \cite{wilhelmi2022end} and \cite{afraz2023blockchain} examine latency and cost, yet their setups often rely on private or controlled environments rather than open testnets—potentially missing on-chain complexities like dynamic gas prices, mempool congestion, and finality scheduling under PoS.

Research on gas-consumption analysis in Ethereum likewise illustrates a partial view. Static analysis tools presented in \cite{gasol2019, runningonfumes} help estimate potential out-of-gas errors, while empirical work in \cite{empiricalgas, gasconsumption} leverages on-chain logs to identify common gas-cost patterns. Although these efforts pinpoint where inefficiencies might occur, they rarely delve into domain-specific architectures featuring multi-contract or multi-administrative logic. Alongside these, gas-reduction guidelines and design patterns (e.g., \cite{gasreduction, gasoptimization, designpatterns, smartmoneywasting}) propose optimizations like minimizing redundant writes or flattening data structures, yet typically validate these strategies in smaller or generic DApps rather than in high-concurrency telecom scenarios.

In tandem, transaction-latency and block-level analyses (e.g., \cite{steponthegas, transactiontimes}) explore how dynamic gas pricing or block-size adjustments can improve inclusion times, but usually treat contracts in an aggregate sense. These works rarely dissect how interdependent calls—such as chaining a breach registration (registerBreach) immediately before penalty calculation (calculatePenalty)—increase costs or delays during periods of network congestion. Furthermore, the literature on PoS concurrency often highlights how rising transaction volumes impact throughput and block propagation but does not specifically address how specialized, multi-role DApps might exacerbate mempool competition or finalization bottlenecks.
Collectively, these studies underscore blockchain’s promise for trusted interactions in telecommunications and beyond, yet four key challenges remain largely underexplored: (i) While standardization bodies outline conceptual frameworks for blockchain in telecom, few provide detailed guidance on implementing complete, multi-contract architectures that enforce role-based logic, handle concurrency, and operate under real-world blockchain conditions. (ii) Existing gas analyses seldom scrutinize the function-by-function costs of domain-specific DApps (e.g., provider registration, SLA breach recording), nor do they consider how concurrency and repeated writes affect resource consumption. (iii) Research on transaction latency and PoS block parameters often overlooks the interplay between frequent on-chain updates, multi-step calls, and concurrency surges in a single DApp—factors that can significantly prolong confirmation times and lastly (iv) although many works propose general best practices, they rarely quantify the real trade-offs or highlight concrete “lessons learned” when multiple administrative domains interact intensively under live testnet conditions.
    
Therefore, in this way our work builds upon ETSI frameworks \cite{ETSI2022SLA,ETSI2024Settle,ETSI2020Applications}, particularly the scenario proposed in \cite{ETSI2024Marketplace}, which envisions a network slicing resource-sharing marketplace. In this scenario, service providers and consumers interact in a trustless environment to exchange virtual network slices on shared physical infrastructure. Inspired by these standards, we develop a set of six smart contracts to enforce role-based access, enable SLA monitoring, and streamline penalty execution, thus automating various facets of inter-provider agreements. Specifically, our approach partitions the service lifecycle into multiple stages—from initial marketplace registration to breach handling and penalty settlement—allowing transparent inter-provider collaboration without centralized oversight. Our work offers a detailed performance evaluation for each contract function, placing particular emphasis on how batch sizes ranging from 2 to 50 (and up to 100) influence gas consumption, transaction finalization times, and network resource utilization.
Our work bridges these gaps by designing and evaluating a domain-specific inter-provider DApp on Ethereum’s Sepolia network. We develop and empirically measure a modular framework that automates role-based interactions—from registration to penalty enforcement—capturing gas consumption at the function level, latency under varying concurrency, and block-level throughput data (e.g., block size, mempool behavior). By examining how each contract’s execution scales with user volume, we offer lessons learned that reconcile known optimization methods (e.g., threshold-based breach logging) with the reality of multi-step, multi-administrative workflows. In doing so, we extend the current literature beyond high-level architectures or generic performance guidelines, demonstrating how specialized, multi-contract logic behaves in truly decentralized, load-intensive contexts.

\section{Background}
\label{sec3}
The 5Growth project \cite{5GAdvance,20205growth} is an EU Horizon 2020 initiative that aims to enhance 5G networks for industries such as Industry 4.0, transportation, and energy by delivering an AI-driven, automated, and shareable end-to-end solution. The project uses network slicing to allow multiple virtual networks to operate on the same physical infrastructure and meet different industry requirements. A key aspect of the project is the development of multi-administrative domains, which permits the management and orchestration of 5G network services across different administrative regions and jurisdictions, ensuring service delivery over varied areas. Building on this framework, an inter-provider agreements system \cite{javed2022blockchainPIMRC,javed2022blockchainglobecom} has been introduced that allows each consumer domain to access specialized capabilities while each provider domain uses idle resources. This system defines the terms of collaboration to prevent over-provisioning and support service delivery, a concept further illustrated by ETSI’s Network Functions Virtualization (NFV) framework \cite{ETSINFV002} that promotes standardization and interoperability among partners. The development of these agreements also emphasizes the need for trust among interacting entities, as transparent service access, consistent capability advertisement, and SLA enforcement depend on verifiable accountability measures. The formation of these agreements involves processes such as negotiation and lifecycle SLA management, which require mechanisms for recording interactions, tracking breaches, and applying penalties to align incentives and prevent opportunistic behavior, thereby supporting the federation of services in next-generation networks.

\subsection{Ethereum Blockchain and Key Network Parameters: An Overview}

A \textit{blockchain} is recognized as a decentralized ledger that permanently documents all activities within its network. In this paper, we exploit and analyze such feature in Ethereum and the Sepolia testnet. This process involves assembling blocks that encompass a batch of transactions, each marked by unique IDs \cite{vujivcic2018blockchain, wood2014ethereum}. After a block is appended to the chain, any attempt to alter its data necessitates rewriting all subsequent blocks. This architectural design ensures that, once enough blocks are appended, the data in any specific block becomes immutable, thus securing the reliability of recorded information \cite{buterin2014ethereum}.

\textit{Smart contracts}, which are programmable applications running autonomously on blockchain platforms for example, Ethereum \cite{EthereumSmartContracts}, facilitate the automatic execution of complex agreements. These contracts, often developed in high-level programming languages such as Solidity \cite{ETSI2021SmartContracts}, replicate the logic and stipulations typically found in traditional contracts, thereby simplifying compliance and governance.

Building on the discussion of blockchain and smart contracts, it is necessary to examine the EVM and PoS mechanism. These systems govern contract execution and block addition, forming the basis for understanding subsequent smart contract operations.

\subsubsection{Ethereum Virtual Machine (EVM): An Overview}

The EVM is the execution environment for smart contracts on the Ethereum blockchain. It ensures that contract code runs deterministically across all nodes while managing computational resources through the use of “\textit{gas}.” Gas is a fundamental concept that quantifies the computational effort required to execute operations, making it central to understanding the costs associated with various smart contract functions such as role assignment, array manipulation, and storage updates \cite{ethereumGas}.

At its core, the EVM maintains a transient \textit{machine state} during transaction execution. This state includes elements such as memory and storage, which are used differently in the context of smart contract operations. Memory is a temporary, byte-addressable space used for intermediate computations, while \textit{storage} is a persistent data area that holds contract state across transactions. The distinction between these two is critical because storage operations are significantly more gas-intensive, especially when they involve writing data to previously uninitialized storage slots. This process, often referred to as “cold” access, requires a higher gas cost compared to “warm” access—when the storage has been previously initialized or updated. One particularly important aspect of gas consumption arises from how the EVM manages state changes. 
Additionally, the EVM enforces gas limits for each transaction. Each operation, including arithmetic, data retrieval (via SLOAD), and data writing (via SSTORE), has a specific gas cost. For instance, the gas cost for writing a non-zero value to storage is higher because it requires altering the state trie \cite{wood2014ethereum}. 

\subsubsection{Proof of Stake (PoS) in Ethereum: An Overview}

Consensus is the general agreement reached among network participants. In the Ethereum blockchain, consensus is achieved when at least 66\% of nodes agree on the network’s global state. The consensus mechanism encompasses the protocols, incentives, and processes that enable nodes to agree on the blockchain's state.  Ethereum now operates under a PoS consensus mechanism, which differs conceptually from the earlier Proof of Work (PoW) model. In PoS, validators secure the network by staking ETH rather than expending computational power to solve cryptographic puzzles. As a result, block finality is achieved more predictably, with stable block intervals and a built-in base fee mechanism that adjusts dynamically with network activity \cite{croman2016scaling,kiayias2017ouroboros,javed2025empiricalsmartcontractslatency}.

Under PoS, when a transaction is submitted, it is first collected in validators’ mempools, similar to PoW. However, a validator’s eligibility to propose a block is determined both by the amount of ETH staked and a randomized selection process, rather than by solving complex puzzles. Once selected, the validator aggregates transactions, executes them through the EVM, and proposes a new block for validation. Other validators then verify the block, and consensus is reached, thereby adding the block to the canonical chain \cite{eyal2016bitcoinng,sai2021performance}.

\subsubsection{Relevant Blockchain Network Parameters}
\label{parameters_2}
As mentioned above transaction on the Sepolia testnet is a digitally signed instruction from an externally owned account. After submission, these transactions are validated and added to the blockchain through consensus process. Several critical network parameters influence this process on Sepolia which are explained as follow: 
\begin{itemize}
    \item Gas Usage: The term \emph{Gas Usage} quantifies the computational effort required to execute a transaction on the EVM. Each opcode—such as storage operations, arithmetic, and event emissions—has a predefined gas cost~\cite{EthereumVirtualMachine}. A simple ETH transfer requires \(21{,}000\,\text{gas}\), whereas \emph{contract transactions} incur higher costs due to complex operations like function calls, state updates, and event logging. When an Externally Owned Account (EOA) triggers a contract, the EVM executes its logic, leading to variable gas costs~\cite{chen2017contracts}. Additionally, the \emph{gas limit} defines the maximum gas per block, safeguarding the network against spam and denial-of-service attacks.
    \item Gas Price: In Ethereum, the \emph{Gas Price}, expressed in Gwei (where \(1\ \text{Gwei} = 10^{-9}\,\text{ETH}\)), represents the cost per unit of computational effort required for executing transactions and smart contracts on PoS-based networks \cite{buterin2019eip1559,javed2025empiricalsmartcontractslatency}. Users essentially \emph{bid} on gas prices to accelerate transaction inclusion in the next block. Validators, who propose and finalize blocks, prioritize transactions with higher fees to maximize revenue, considering the fee-burning mechanism introduced by EIP-1559. Gas prices fluctuate based on network congestion—rising when demand increases and falling during low activity—effectively regulating computational workload. The \emph{mempool} acts as a temporary storage for pending transactions, where \emph{mempool time} denotes the waiting period before a transaction is confirmed and added to a block.
    \item Block Size: In Ethereum, a block serves as a container for transactions, with its \emph{block size} (measured in bytes) indicating the total data volume of those transactions. This size affects network latency and throughput, as larger blocks require more data to be propagated across nodes. Ethereum enforces a \emph{block gas limit}, typically around 30 million gas, which restricts the total computational effort per block. Since gas measures computational cost rather than byte size, the number and complexity of transactions influence the actual block size. A block filled with fewer but complex transactions may be smaller in bytes than one with many simpler transactions. 
    \item Transaction Count: \emph{Transaction Count} refers to the total number of transactions recorded within a single Ethereum block, serving as a key measure of the network’s throughput. This count is influenced by the \emph{block gas limit} and the \emph{gas usage} per transaction. Simple transactions, such as basic Ether (\text{ETH}) transfers, consume minimal gas, allowing more to fit within a block. Conversely, complex transactions involving smart contracts, multiple function calls, or extensive state changes require more gas, limiting the number of transactions per block. As a result, blocks with higher computational complexity reach the gas limit with fewer transactions, affecting overall network performance by determining how many transactions can be processed within each block interval.
\end{itemize}


\begin{figure}
    \centering
    \includegraphics[width=1\columnwidth]{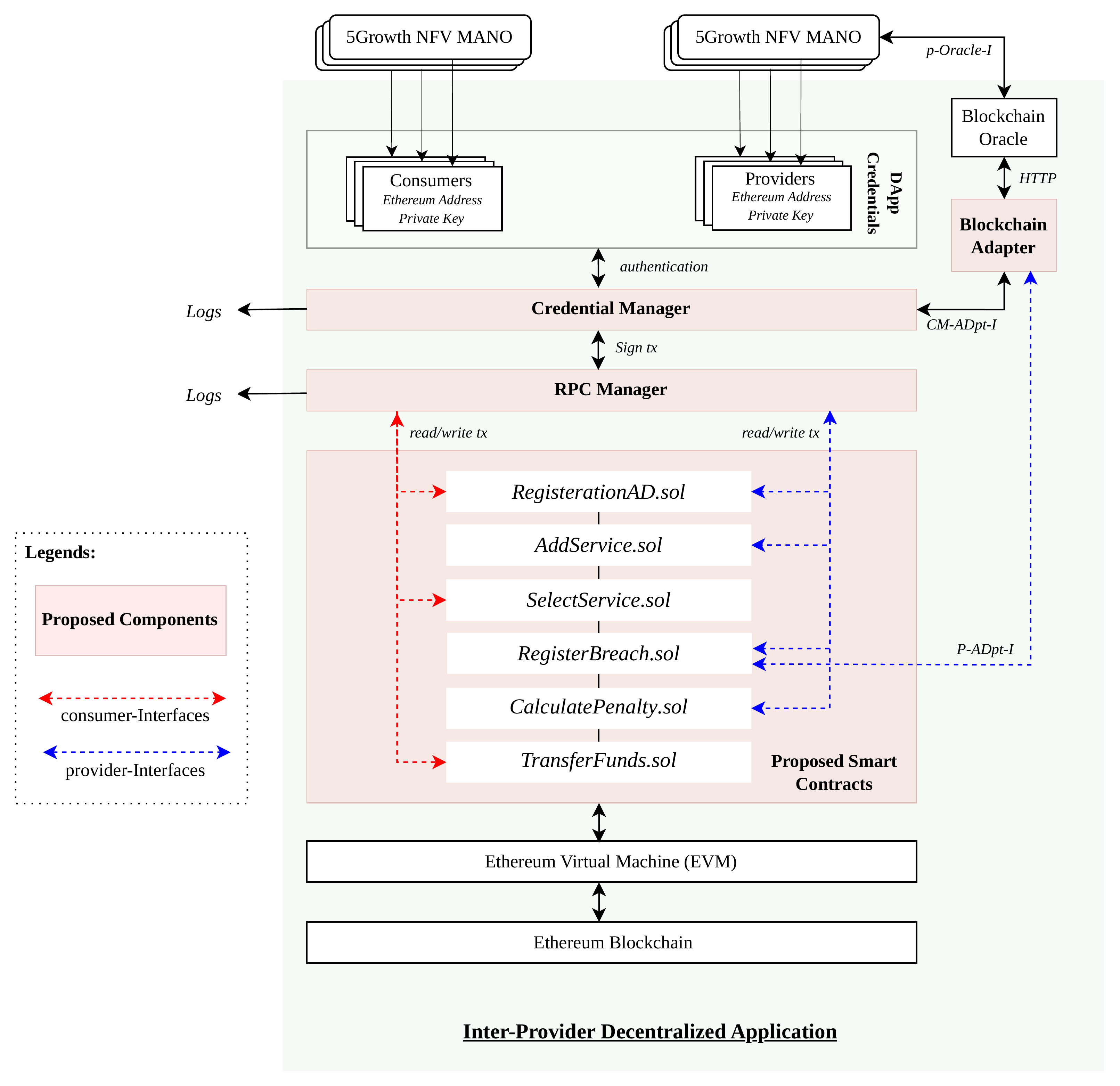}
    \caption{Inter-Provider DApp Architecture: Key components include the Credential Manager, RPC Manager, Blockchain Adapter, and multiple smart contracts deployed on the Ethereum blockchain. Consumer and provider interfaces enforce role-based access control for proposed smart contract interactions.}
    \label{Framework}
\end{figure}

\section{Proposed Framework: Decentralized Application (DApp) on Ethereum Blockchain for 6G Inter-Provider Agreements}
\label{sec4}
In this Section, we explain the proposed components and interfaces as highlighted in Figure \ref{Framework} in our DApp to facilitate inter-provider agreements. 

\subsection{System Overview}
Figure \ref{Framework} presents the overall architecture of the proposed multi-contract Inter-Provider DApp for enabling inter-provider agreements, showing two Administrative Domains—labeled “Consumer” and “Provider”—that align with the 5Growth reference architecture. Within each domain, core functional blocks, such as the SLA Manager, AI/ML Platform, and Monitoring System \cite{20205growth}, provide domain-specific capabilities. 

The Inter-Provider dApp architecture connects these domains with an on-chain environment through a set of components that coordinate interactions and uphold secure operations. The \textit{Credential Manager} generates and manages cryptographic keys and handles transaction signing to support contract deployment and interaction, a process kept separate from higher-level modules yet ensuring consistent security. Through the \textit{CM-ADpt-I} interface, the Credential Manager interacts with the \textit{Blockchain Adapter}, which exchanges data between off-chain services and the blockchain. The \textit{RPC Manager} forwards transaction requests and responses, relying on Remote Procedure Calls (RPC) to link the smart contracts to the blockchain and thereby hide its complexity. The \textit{p-Oracle-I} interface ties external oracle services to the Blockchain Adapter, feeding off-chain data into smart contracts and supporting more precise state updates. This arrangement also provides consumer and provider interfaces that govern access to resource-sharing functions according to assigned roles, so that only authorized entities can trigger smart contract actions and track agreements transparently.

Building on this foundational framework, the Inter-Provider DApp incorporates multiple smart contracts to facilitate the complex interactions between different providers and consumers. Smart contracts such as \contract{RegistrationAD.sol}, \contract{AddService.sol}, \contract{SelectService.sol}, \contract{RegisterBreach.sol}, \contract{CalculatePenalty.sol}, and \contract{TransferFunds.sol} each encapsulate distinct functionalities ranging from Administrative Domain (AD) registration to service offering, selection, penalty assessment, and fund transfers. Interaction with these smart contracts is managed through the Smart Contract Interface, which is highlighted as \textit{consumer-Interfaces} and \textit{provider-Interfaces}. Both Consumer and Provider domains interact with the DApp through these proposed interfaces to ensure seamless communication. On the provider side, an additional functionality retrieves performance or KPI data from an external Blockchain Oracle, ensuring that off-chain data such as SLA metrics is securely and verifiably brought on-chain. This integration is facilitated by the \textit{p-ADpt-I} interface, which connects the provider’s off-chain monitoring systems with the on-chain environment. In doing so, the adapter integrates with the smart contracts to obtain performance metrics and passes them to \contract{RegisterBreach.sol}, thereby enabling automated SLA breach detection and penalty enforcement.

Finally, the underlying Blockchain Network Ethereum Blockchain serves as the trusted execution layer for all smart contracts. 

\subsection{Proposed Smart Contracts}
\label{smart-contracts}
Central to our framework are smart contracts designed to facilitate inter-provider agreements. Below, we explain each of these smart contracts presented in Figure \ref{Framework}.

\begin{itemize}
    \item \contract{RegistrationAD.sol}: The initial step in the inter-provider agreements DApp involves authenticating Administrative Domains (ADs). This process ensures that each AD is uniquely identifiable and authorized to participate in subsequent service exchanges, thereby upholding the integrity of the overall framework. In the proposed system, an AD begins its participation by calling the \function{registerAD} function in the \contract{RegistrationAD.sol} smart contract, passing a specified role as follow:
    \noindent
    \begin{center}
    \begin{minipage}[t]{0.44\textwidth}
    \vspace{0pt}
    \begin{lstlisting}[language=Solidity, 
    basicstyle=\ttfamily\scriptsize,  % Smaller font
    columns=fullflexible,             % More flexible column width
    frame=none, 
    numbers=left, 
    firstnumber=1, 
    breaklines=true]
    contract RegistrationAD {
    enum Role { Consumer, Provider }

    mapping(address => AD) public ADs;
    uint256 public ADCount;

    struct AD {
        address ADAddress;
        uint256 registrationTime;
        Role role;
    }
    \end{lstlisting}
    \end{minipage}%
    \hspace{-3mm}
    \begin{minipage}[t]{0.44\textwidth}
    \vspace{0pt}
    \begin{lstlisting}[language=Solidity, 
        basicstyle=\ttfamily\scriptsize, 
        columns=fullflexible, 
        frame=none, 
        numbers=left, 
        firstnumber=12, 
        breaklines=true]
    function registerAD(Role role) public {
        bool isNewAD = (ADs[msg.sender].ADAddress == address(0));
        ADs[msg.sender] = AD(msg.sender, block.timestamp, role);
        
        if (isNewAD) {
            ADCount += 1;
        }
    }
    \end{lstlisting}
    \end{minipage}
    \end{center}

    The above shown contract declares a Solidity enum named \texttt{Role} to distinguish between two key roles: \textit{Consumer} and \textit{Provider}. This choice of using an enum provides type safety and role clarity, ensuring that each AD can only be assigned one of these well-defined categories. In a multi-domain 5G or beyond-5G environment—where providers offer services and consumers utilize those services—strictly separating roles is crucial for enforcing service-level boundaries and mitigating potential conflicts or ambiguous privileges.

    Upon invocation of \function{registerAD} with the desired role, the contract records the caller’s Ethereum address on-chain and associates it with the chosen \texttt{Role}, along with a timestamp reflecting the moment of registration. Internally, the contract tracks all registered ADs via the \texttt{ADs} mapping and maintains a counter \texttt{ADCount} to monitor the total number of enrolled ADs. If the address has not previously registered (\textit{isNewAD}), the counter is incremented to reflect the new entrant.

    This role-based architecture directly governs how each AD interacts with the rest of the DApp. Specifically, an AD registered as a \textit{Provider} is granted the authority to list and manage services on-chain, while an AD marked as a \textit{Consumer} is restricted to discovering and procuring those services. By setting these permissions rigorously at registration, the framework ensures that participants can only invoke smart contracts that align with their designated roles. This not only mitigates unauthorized operations but also streamlines subsequent processes such as service discovery, selection, and penalty enforcement—all of which hinge on the Consumer–Provider relationship established at registration.
    \item \contract{AddService.sol}: After an AD has successfully registered as a \textit{Provider}, the next step is adding services to the decentralized framework. The \contract{AddService} smart contract defines the \function{addService} procedure, which allows providers to register their service offerings on-chain. When a provider calls \function{addService}, its Ethereum address (\textit{msg.sender}) is captured along with three parameters—\textit{serviceId}, \textit{location}, and \textit{cost}—that describe the new service. This design choice reflects the principle of least privilege in an inter-provider framework, ensuring that only addresses registered as \textit{Provider} in the DApp’s role-based model can invoke this function.
    Service information is managed within a dedicated \texttt{Service} struct, which encapsulates essential attributes, including the provider’s address, a unique service identifier, and the cost of the service. Instead of storing services individually, they are recorded in a global \texttt{services} array, while a \texttt{mapping} associates providers with their registered service indices. The \function{addService} function is implemented as follows:
    \begin{center}
    \begin{minipage}{0.85\textwidth}
    \vspace{0mm}
    \begin{lstlisting}[language=Solidity, basicstyle=\ttfamily\small, columns=fixed, frame=none, numbers=left, firstnumber=1, breaklines=true]
    mapping(address => uint256[]) public providerServices;

    function addService(string memory serviceId, string memory location, uint256 cost) public {
    services.push(Service(msg.sender, serviceId, location, cost));
    providerServices[msg.sender].push(services.length - 1);
    }
    \end{lstlisting}
    \vspace{-3mm}
    \end{minipage}
    \end{center}
    By organizing service information in a structured and immutable manner, this contract enables transparent registration and retrieval of provider services. The use of mappings optimizes lookup efficiency, ensuring that each provider can quickly access their registered services without iterating through the entire dataset.
    \item \contract{SelectService.sol}: Once one or more services are added, consumers can discover and choose among these offerings. The \contract{SelectService} contract provides the \function{serviceSelection} procedure, marking the next logical step in the inter-provider agreement lifecycle. When a Consumer calls \function{serviceSelection}, the contract first verifies that the specified provider exists and has registered services. This is achieved by retrieving the provider’s record and ensuring the associated \texttt{providerAddress} is valid. 

    Next, the contract checks whether the requested service is valid by referencing the provider’s list of registered services and ensuring that the service ID is correctly mapped. If both conditions are met, the contract records the selection by appending a \texttt{Selection} struct to the contract’s \texttt{selections} array. This establishes an immutable link between the consumer’s Ethereum address, the chosen provider, and the selected service. The final step is emitting a \textit{ServiceSelected} event, ensuring that the selection process is auditable and that subsequent interactions—such as payment processing or SLA enforcement—can reference this transaction. The function is implemented as follows:

    \begin{center}
    \begin{minipage}{0.85\textwidth}
    \vspace{0mm}
    \begin{lstlisting}[language=Solidity, basicstyle=\ttfamily\small, columns=fixed, frame=none, numbers=left, firstnumber=1, breaklines=true]
    function serviceSelection(address provider, uint256 serviceId) public {
    Provider storage p = providers[provider];
    require(p.providerAddress != address(0), "Provider not found");

    Service storage service = p.services[serviceId];
    require(service.serviceId != 0, "Service not found");

    selections.push(Selection(msg.sender, provider, serviceId));
    emit ServiceSelected(msg.sender, provider, serviceId);
    }
    \end{lstlisting}
    \vspace{-3mm}
    \end{minipage}
    \end{center}

    By enforcing role-based access, this function ensures that only consumers can initiate service selection while preventing unauthorized modifications to provider records. The modular separation of \contract{AddService} and \contract{SelectService} enables clear, permissioned interfaces for both providers and consumers, reducing attack surfaces and facilitating transparent auditing in inter-provider agreements.

    \item \contract{RegisterBreach.sol}: This contract provides the \function{registerBreach} function, which records SLA breaches for each provider using the \texttt{breachCount} mapping. Breach detection relies on KPIs fetched via the Blockchain Adapter using a Blockchain Oracle. When a provider calls \function{registerBreach} with a specified number of breaches, the function updates the provider’s total breach count. To maintain an auditable on-chain record, a \texttt{BreachRegistered} event is emitted. After recording the breach, the function invokes \function{calculatePenalty} in the \contract{CalculatePenalty} contract to process penalty assessment separately. This separation of concerns ensures that penalty computation is handled in a dedicated module, reducing complexity in breach tracking. The function is implemented as follows:

    \begin{center}
    \begin{minipage}{0.85\textwidth}
    \vspace{0mm}
    \begin{lstlisting}[language=Solidity, basicstyle=\ttfamily\small, columns=fixed, frame=none, numbers=left, firstnumber=1, breaklines=true]
    function registerBreach(uint256 numBreaches) public {
    breachCount[msg.sender] += numBreaches;
    emit BreachRegistered(msg.sender, numBreaches);
    calculatePenalty(msg.sender);
    }
    \end{lstlisting}
    \vspace{-3mm}
    \end{minipage}
    \end{center}

    \item \contract{CalculatePenalty.sol}: The \function{calculatePenalty} procedure determines the penalty imposed on a provider after an SLA breach is recorded. The penalty follows a linear model, computed as \( \text{penalty} = \text{fidelityFee} \times \text{breachCount} \), where \texttt{fidelityFee} is a predefined constant that establishes the per-breach penalty rate, and \texttt{breachCount} represents the total number of infractions committed by the provider. Once calculated, the penalty is stored in the \texttt{penalties} mapping, linking each provider to their respective financial penalty. The contract then emits a \texttt{PenaltyCalculated} event to ensure transparency in penalty enforcement. The function is defined as follows:
    \begin{center}
    \begin{minipage}{0.85\textwidth}
    \vspace{0mm}
    \begin{lstlisting}[language=Solidity, basicstyle=\ttfamily\small, columns=fixed, frame=none, numbers=left, firstnumber=1, breaklines=true]
function calculatePenalty(address provider) public {
    uint256 penalty = fidelityFee * breachCount[provider];
    penalties[provider] = penalty;
    emit PenaltyCalculated(provider, penalty);
    }
    \end{lstlisting}
    \vspace{-3mm}
    \end{minipage}
    \end{center}

    By structuring breach registration and penalty assessment as distinct operations, this model ensures modularity, enabling future modifications such as adjustable penalty rates, automated settlements, or integration with on-chain dispute resolution mechanisms.

    \item \contract{TransferFunds.sol}: Once the breach or penalty has been registered and recorded, the final step in settling inter-provider obligations is to transfer the appropriate funds from the consumer’s wallet to the provider’s wallet. In the proposed framework, this is facilitated by a dedicated \contract{TransferFunds} contract, which ensures direct, on-chain compensation without relying on intermediaries. The contract’s core function, \function{transferFunds}, takes two essential inputs: the \texttt{recipient} address (defined as \textit{payable}) and the amount of Ether sent by the caller in \textit{msg.value}. Before executing the transfer, the function verifies that \textit{msg.value} is nonzero, ensuring a valid and nonempty payment. Once satisfied \texttt{recipient.transfer(msg.value)} immediately sends the specified Ether to the designated provider’s address, finalizing the compensation sequence. This design preserves transparency: on-chain records show the exact transfer amounts, and only the contract’s authorized logic can invoke fund transfers. By embedding this functionality in a distinct contract, the framework enforces role-based control—only a consumer is permitted to pay a provider—and retains a clear audit trail of all financial settlements. In multi-domain settings, where multiple providers and consumers interact, eliminating external arbitration and manual reconciliation is an important advantage. The \contract{TransferFunds} contract prevents disputes over payment amounts or schedules by codifying the settlement process on-chain \cite{ETSI2024Settle}.

\end{itemize}

\begin{figure}[t!]
    \centering
\includegraphics[width=1\linewidth]{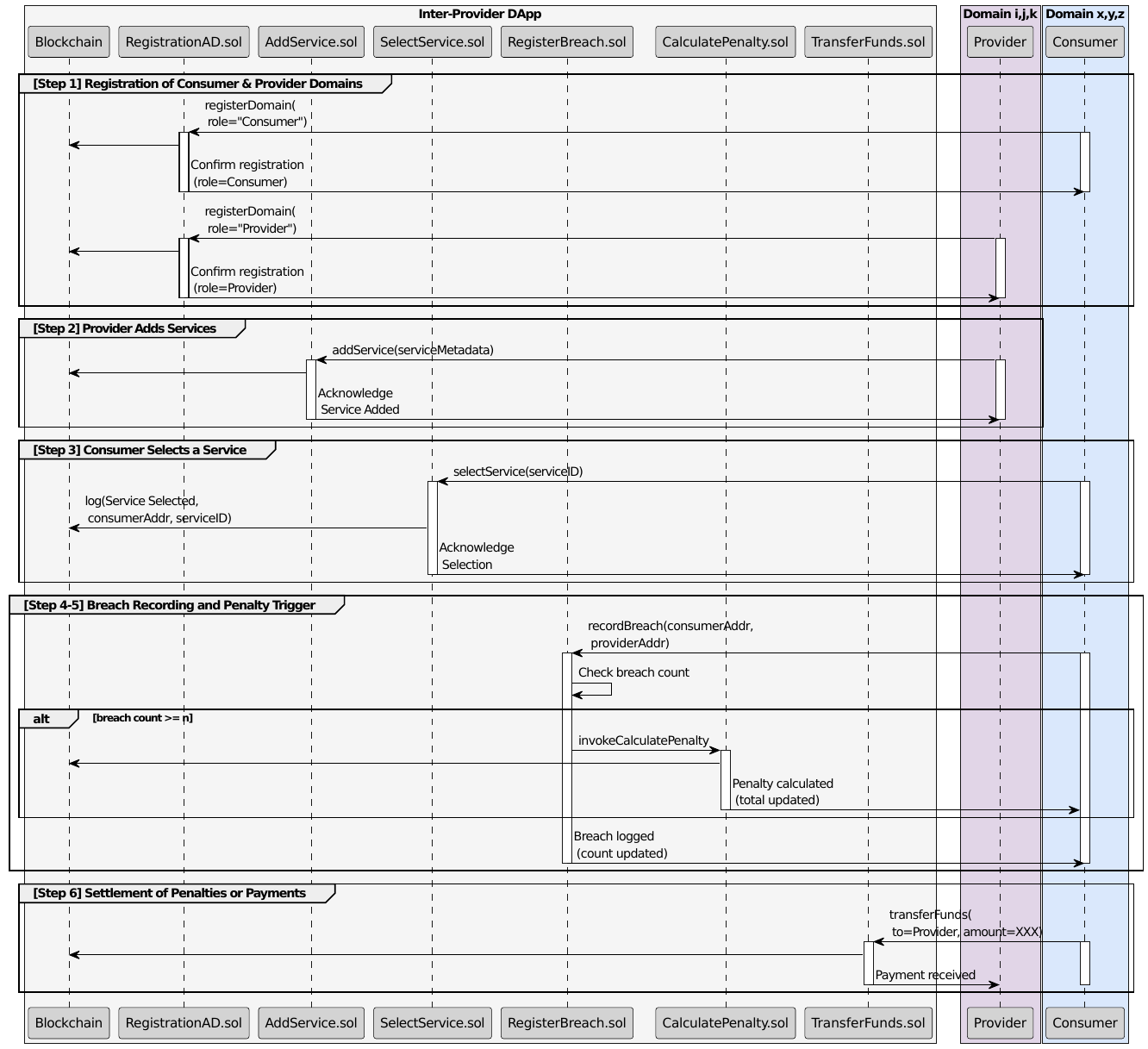}
   \caption{Sequence Diagram: Illustrating the step-by-step interactions between providers, consumers, and smart contracts within the inter-provider DApp.}
    \label{sequece}
\end{figure}

\subsection{Workflow}
\label{workflow}
In the proposed framework, Figure \ref{sequece} illustrates a comprehensive workflow that seamlessly integrates on-chain smart contract interactions with critical supporting components. The process begins with the registration of consumer and provider domains ([Step 1]), where the \contract{RegistrationAD.sol} contract records each domain’s Ethereum address and assigns it the appropriate role. At the same time, the \textit{Credential Manager} ensures that only authorized entities gain access, establishing a secure foundation for all subsequent interactions.

Once registration is complete, providers are empowered to add new services through the addService function in \contract{AddService.sol }([Step 2]). This step not only updates an on-chain catalog of available services but also benefits from the \textit{RPC Manager}, which channels transaction requests to the EVM, thereby ensuring a transparent and reliable update of service records. Following service addition, consumers proceed to select a service using the serviceSelection function in \contract{SelectService.sol} ([Step 3]). This selection creates a permanent link between the consumer, the provider, and the chosen service, setting the stage for later transactions and SLA monitoring. Behind the scenes, the \textit{Blockchain Adapter} facilitates smooth data interchange between off-chain interactions and on-chain records, reinforcing the system's integrity.

As the DApp continues its operation, it monitors service performance through integrated oracles. In cases where a breach of the service-level agreement occurs, consumers can record these incidents via the \contract{RegisterBreach.sol} contract using parameters from Blockchain Adapter and integrated smart contract. When the number of breaches reaches a pre-determined threshold—typically three—the \contract{CalculatePenalty.sol} contract is automatically invoked to compute the appropriate fine ([Step 4-5]). This process is supported by external interfaces that verify real-time performance data, ensuring that the penalty calculations are both accurate and timely. The final phase of the workflow involves the settlement of penalties or service charges. Through the \function{transferFunds} function in \contract{TransferFunds.sol} ([Step 6]), funds are securely transferred directly on the blockchain from consumer to provider, eliminating any risk of unauthorized transactions.


\section{Deployment and Evaluation}
\label{sec5}
In this Section, we assess the performance of our blockchain framework for inter-provider agreements. The evaluation includes benchmark tests on a system running Kubuntu 22.04.3 LTS with 32 GB of DDR4 RAM and 16 GB of disk storage. This system simulates operational conditions for blockchain applications.

The Hardhat development framework is used for compiling, deploying, and testing Ethereum smart contracts. Its integration with Solidity and built-in testing tools support the development and debugging processes. Similarly, the Sepolia testnet is selected because it replicates network conditions and parameters similar to those of the Ethereum mainnet while allowing controlled testing. This testnet supports the execution of smart contracts in an environment that reflects live blockchain operations. Also, Alchemy is employed as the node service provider to ensure connectivity with the blockchain network and to provide gas estimation services. Its use facilitates reliable interactions during the testing phase.

Lastly, the simulation uses 100 Externally Owned Accounts (EOA) with associated private keys to represent multiple user entities. The number of accounts is chosen to allow the execution of transactions from a range of sources and to evaluate the scalability of the framework. The testing procedure is conducted over 10 iteration rounds.

\begin{table}[t]
\centering
\small
\begin{adjustbox}{max width=\columnwidth}
\begin{threeparttable}
\caption{Simulation system setup for inter-provider agreements on blockchain}
\label{tab:system_setup}
    \begin{tabular}{p{7cm} p{8cm}} 
        \toprule
        \textbf{Component} & \textbf{Specification} \\ \toprule
        Operating system & Kubuntu 22.04.3 LTS\\ \midrule
        Memory (RAM) & 32 GB DDR4 \\ \midrule
        Blockchain testnet & \href{https://sepolia.etherscan.io/}{Sepolia testnet} \\ \midrule
        Development framework & \href{https://hardhat.org/}{Hardhat} 2.22.4 \\ \midrule
        Smart contract language & \href{https://soliditylang.org/}{Solidity} v0.8.0 \\ \midrule
        Number of accounts & 100 Externally Owned Accounts (EOA) and their private keys \\ \midrule
        Node service & \href{https://www.alchemy.com/}{Alchemy} \\ \midrule
        Iteration Rounds & 10 \\ \bottomrule
    \end{tabular}
    \vspace{5pt}
    \begin{minipage}{0.95\columnwidth}
        \small \textit{Note}: \textit{The \textbf{Sepolia} testnet is used as the default blockchain environment for live blockchain parameters and realistic environment. \textbf{Hardhat} serves as default the Ethereum development framework, with \textbf{Solidity} as the smart contract programming language. The setup includes 100 EOAs with private keys, and \textbf{Alchemy} is used as the node service provider for reliable connectivity and gas estimation. The simulation runs for 10 iteration rounds.}
    \end{minipage}
\end{threeparttable}
\end{adjustbox}
\end{table}

\begin{figure}
    \centering
    \includegraphics[width=1\columnwidth]{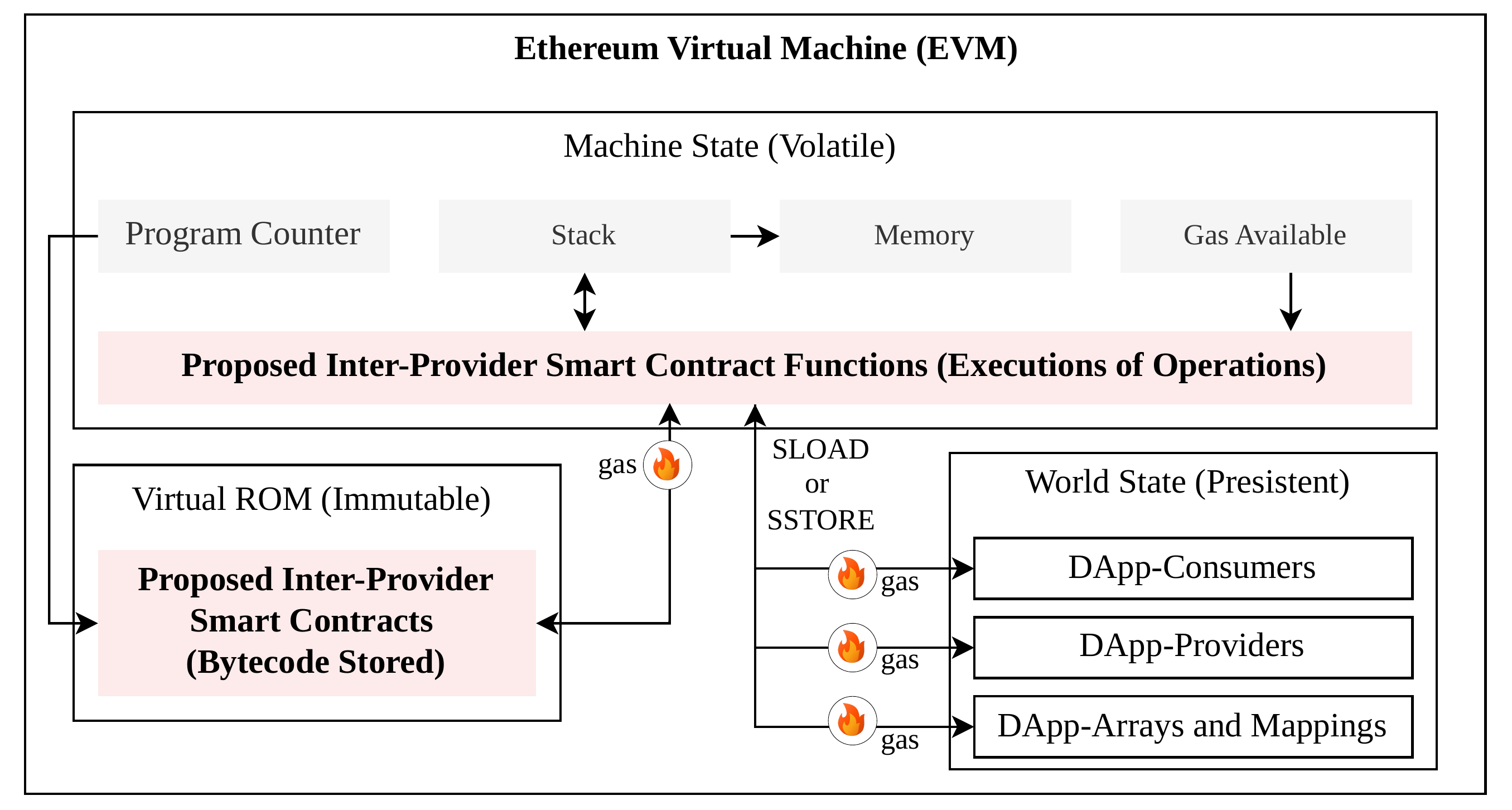}
    \caption{EVM Execution Flow and Gas Consumption: Interaction between proposed smart contract execution, volatile machine state, immutable bytecode storage, and persistent world state, highlighting gas costs for storage operations (SLOAD/SSTORE).}
    \label{EVM-contracts}
\end{figure}

We begin our evaluation by examining the smart contracts introduced in Section~\ref{smart-contracts}, which are deployed on the Sepolia testnet. These contracts, together with their associated functions, are summarized in Table~\ref{tab:smart_contracts}. They form the core of our inter-provider agreements and operate in multiple phases (creation, monitoring, and billing) on the Ethereum blockchain. Figure~\ref{EVM-contracts} illustrates the execution environment within the Ethereum Virtual Machine (EVM), showing how these contracts manage state transitions, data storage, and computational tasks.

In assessing the performance of these contracts, two metrics are particularly important: (i) \textit{Gas Consumption:} A measure of the computational resources required to execute a transaction. High gas consumption often indicates complex or storage-intensive operations. (ii) \textit{Execution time:} The time from when a transaction is submitted to when it is confirmed on the blockchain. This metric reflects the responsiveness of the network and any delays related to block production or mempool congestion.

In addition to these metrics, we provide block-level details such as the Gas Price (Gwei), Block Size (KB), the number of transactions within the block, and the measured latency. By analyzing these data points, we offer a detailed perspective on how the network handles contract execution under varying loads, particularly as the number of ADs increases.


\subsubsection{Gas Usage}
\label{define-gas}
Ethereum enables the execution of smart contracts through the EVM. The EVM is a virtual machine that interprets a low-level, stack-based \textit{bytecode} language (see Figure \ref{EVM-contracts}), which allows it to carry out various computational tasks necessary for processing smart contracts. At the heart of the EVM's functionality are \textit{opcodes}, which are fundamental instructions that direct the virtual machine's operations. The smart contract’s immutable bytecode is stored in the EVM’s “Virtual ROM,” while the volatile machine state (e.g., stack, memory, program counter, and available gas) changes dynamically as operations are executed.

Function calls within smart contracts are central to the execution logic, encompassing operations such as computations, data retrieval, and conditional checks. When a function is called, it may invoke a series of opcodes depending on the nature of the operation it performs. These opcodes can either read from or write to the blockchain's persistent \textit{world state}, influencing how the contract behaves and interacts with data. One of the key opcodes associated with function calls is \textit{SSTORE} (Storage Store), which is used to store data in the blockchain’s state. This opcode is critical because it writes data to the Ethereum ledger, making permanent changes to the contract's state. Because of its role in modifying the blockchain, \textit{SSTORE} is considered a gas-intensive operation. The high gas cost reflects the computational effort and the necessity of achieving consensus across the network to validate the change.

As illustrated in Figure \ref{EVM-contracts}, each operation within our proposed inter-provider smart contract functions triggers changes to the volatile machine state (stack, memory, gas), and when necessary, performs reads (\textit{SLOAD}) or writes (\textit{SSTORE}) to the persistent world state (where DApp-Consumers, DApp-Providers, and various data structures like arrays or mappings reside). Every transaction that involves these opcodes requires \textit{gas}, a unit that measures the computational effort needed to execute operations on Ethereum. 

In our tests, upon the successful execution and confirmation of a transaction, we generate a transaction receipt as:
\textit{const gasUsed = receipt.gasUsed;}. This receipt provides a detailed record of the transaction, encapsulating key information such as the execution status and any events that were triggered during the process. A crucial element of this receipt is \textit{gasUsed}, which quantifies the actual amount of gas consumed during the transaction. This metric is vital as it reflects the computational resources expended to facilitate the transaction, offering insights into the operational complexity of the smart contract's functions within our inter-provider DApp framework.

\subsubsection{Execution Latency}
\label{define-latency}

Latency is a critical performance metric in blockchain systems, measuring the time elapsed from when a transaction is submitted to the network to when it is fully confirmed. This metric is used to assess the responsiveness of the network and serves as a key indicator of overall performance, particularly when executing smart contracts.

In our evaluations, we measure latency by calculating the difference between the timestamp of the block containing the confirmed transaction (\(T_{\text{confirmed}}\)) and the timestamp when the transaction was initially submitted (\(T_{\text{submitted}}\)): \(L = T_{\text{confirmed}} - T_{\text{submitted}}\). This calculation provides a direct measure of the total latency, which includes both the time spent in the mempool (waiting to be picked up by a validator) and the time taken for the transaction to be included in a block and confirmed. The Sepolia testnet, which closely mirrors the Ethereum mainnet, is used in this analysis to evaluate transaction processing under test conditions. By using blockchain-native timestamps, this approach allows us to capture the total response time of transactions on the network.

Our analysis also aims to understand the factors influencing latency, including smart contract function design, concurrency, and blockchain network-level parameters. For example, these parameters include gas price (the fee paid per unit of gas), block size (the total number of transactions that can fit within a block), network congestion (the volume of pending transactions), and transaction complexity (such as state-changing transactions).

To explore how these factors influence latency, we perform batch transactions for various functions of the smart contracts, as detailed in Table~\ref{tab:smart_contracts}. By sending multiple interactions from accounts, we can observe how the network handles these transactions simultaneously and how this impacts overall latency. 

\begin{table*}[ht]
\centering
\footnotesize
\renewcommand{\arraystretch}{1.3} 
\setlength{\tabcolsep}{10pt} 
\begin{threeparttable}
\caption{Overview of Proposed Smart Contract, Key Functions and Their Functionalities}
\label{tab:smart_contracts}
\begin{tabular}{p{2.5cm} p{2.8cm} p{7cm}} 
\toprule
\textbf{Smart Contract} & \textbf{Key Function} & \textbf{Functionality} \\ 
\midrule
\contract{RegistrationAD.sol} & 
\function{registerAD} & Registers each Administrative Domain (AD), assigning roles as consumer or provider. \\ \midrule 

\contract{AddService.sol} & 
\function{addService} & Enables providers to add $n$ number of services to the DApp. \\ \midrule 

\contract{SelectService.sol} &  
\function{serviceSelection} & Allows consumers to select a service from a list of available services by different providers. \\ \midrule 

\contract{RegisterBreach.sol} & 
\function{registerBreach} & Logs breach occurrences associated with a provider. \\ \midrule 

\contract{CalculatePenalty.sol} & 
\function{calculatePenalty} & Computes penalties for providers based on recorded breaches. \\ \midrule 

\contract{TransferFunds.sol} & 
\function{transferFunds} & Facilitates the transfer of funds from consumers to providers. \\  
\bottomrule
\end{tabular}
\end{threeparttable}

\end{table*}

\subsection{Complexity-Driven Gas Usage} 

\begin{figure}
    \centering
    \includegraphics[width=0.5\linewidth]{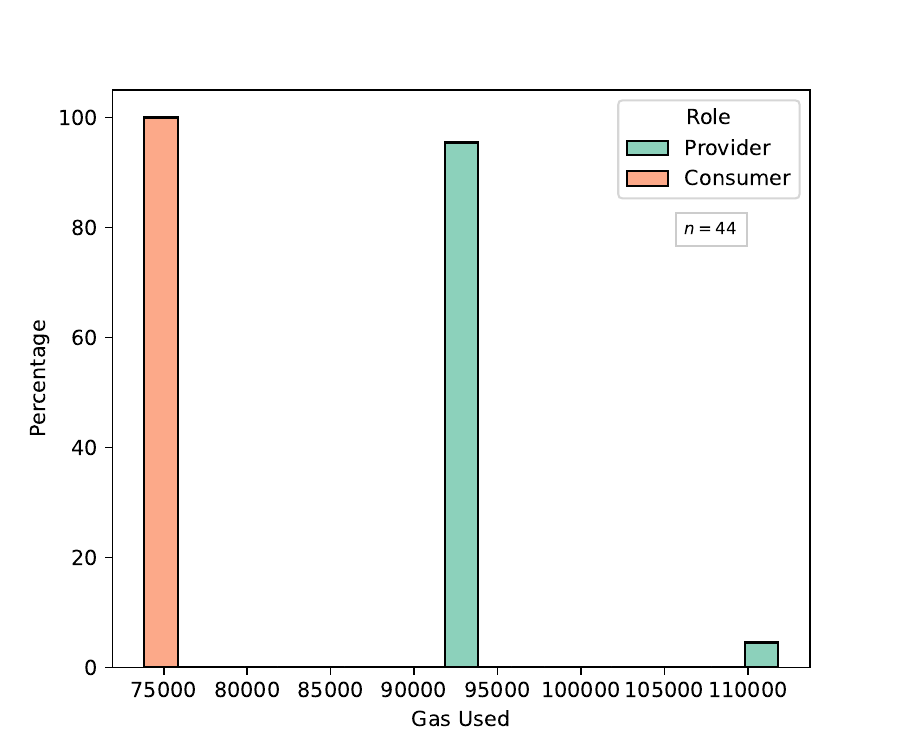}
    \caption{Gas for registration of the ADs for batchsize 44}
    \label{fig:gas-registeration}
\end{figure}

\subsubsection{Registration of an AD}
\label{gas-creation-explaination}

As previously discussed in Ethereum, 'gas' measures the computational effort and cost for transactions, especially 'write operations' that modify blockchain data. Below we provide a comprehensive analysis of gas usage for three integral write operation during creation of an agreement within the smart contract process of agreement creation also highlighted in \ref{tab:smart_contracts}: \function{registerAD}, \function{addService}, and \function{serviceSelection}.

In our DApp, we perform an analysis on a single write operation in the \contract{RegistrationAD.sol} smart contract to register two distinct types of authorized ADs: consumers (denoted as $C_n$) and providers (denoted as $P_n$). Each type of AD is assigned a specific role, either \textit{Consumer} or \textit{Provider}, to enable role-based access control to different functionalities within our DApp. To achieve this, we use an enumerated type (enum) in Solidity, defined in \contract{RegistrationAD.sol} as follows: \textit{enum Role { Consumer, Provider }}. In Solidity, enums are internally treated as integers, with values starting from 0. Thus, \textit{Consumer} is implicitly assigned a value of 0, and \textit{Provider} a value of 1. To analyze registration process, we focus on the gas consumption associated with registering these roles. Specifically, we observed the gas usage for a batch of registrations for different number of user. However, for illustration we select only the batch size 44 (denoted as \textit{ADBatch} with $n = 44$), where 22 users are registered as \textit{Consumer} and 22 as \textit{Provider}.  The histogram of gas used when an account arrives and registers itself as a 'provider' first. The Gas Used for the registration of providers and consumer are illustrated in Figure \ref{fig:gas-registeration}, reveals three distinct values for gas consumption. The Provider role exhibits a slightly more varied pattern. The majority of Provider transactions, specifically 21 out of 22 (95.45\%), used gas within the (90000, 95000] bin, indicating a high level of consistency similar to Consumers but at a higher gas usage level. Additionally, there is a notable outlier where 1 out of 22 Provider transactions (4.55\%) consumed gas in the (105000, 110000] bin. This outlier reflects a deviation from the typical transaction pattern, which shows an initial peak, which is the gas amount needed for the first-time state initialization in Ethereum. For thorough analysis, a consumer was registered first and the gas cost observed was 90927. This shows that the initialization cost is always 17000 in addition to the registration cost and assigning of a role.  \\ State initialization in Ethereum is a key concept that impacts cost of executing smart contracts on the blockchain. This process involves setting a storage slot in a smart contract to a value other than its default (zero) for the first time \cite{jezek2021ethereum}. The state trie is a data structure that combines elements of both a Merkle tree and a Patricia tree  \cite{wackerow2024merkle}. It is designed to optimize the way data is stored and retrieved within the network. When a storage slot of a smart contract is assigned a value for the first time—this process is termed as 'cold' access \cite{ayub2023storage} \cite{wood2014ethereum}. This initial assignment requires modifications to the state trie, which is computationally expensive. For instance, the first registration of a provider in a contract is observed 110,839  due to the complexity of this state initialization. After this 'cold' setup, any subsequent accesses to this storage slot are referred to as 'warm' accesses\footnote{https://www.evm.codes/}.  For example, registering each subsequent provider after the initial one typically costs around 93,739— less than the initial registration. Since in this case, provider gets registered first therefore the initialization gas cost in this case observed is 17000. Similarly in contrast all Consumer transactions consistently utilized gas within the (75000, 80000] bin, with 22 out of 22 transactions (100\%) falling within this range. This is because the enum role assigned 0 for consumer. When registering users in a smart contract, there are distinct differences in gas costs associated with the roles of "Provider" and "Consumer" due to the nature of state updates in the Ethereum blockchain's state trie. 

For example, when registering a 'Provider', the raw transaction input\footnote{https://sepolia.etherscan.io/tx/0x20b6afa10fc5ab10fba622e8bfc1770de14cf67625f97457d2ef9d7d164f37cf} can be observed in hexadecimal format: 0x61d689fa...00000001. This data string can be broken down into two main parts: (i) function selector 0x61d689fa-derived from the first 4 bytes of the Keccak-256 hash of the function signature, this identifies which function to call within the smart contract and the (ii) parameter ...00000001 represents a 256-bit number where the value 1 at the end signifies the role of 'Provider' being assigned in the function call.
Specifically, the role of "Provider" corresponds to a non-zero value in the enumeration (value 1), while the role of "Consumer" corresponds to the default value (value 0). When a user registers as a "Consumer," the smart contract checks the role value to be stored. Since "Consumer" is associated with 0, which is the default value for uninitialized storage, the state trie does not require an update if the storage already implicitly holds this value. In essence, no state change occurs, as the value remains at its default, thus avoiding additional gas costs.  Conversely, registering a user as a "Provider" involves a different process. The role value to be stored is 1, which differs from the default value of 0. Consequently, the smart contract must perform a write operation to the state trie to update the role from 0 to 1. This state trie update is inherently more gas-intensive because it necessitates altering and storing new information on the blockchain. The increased gas consumption associated with registering a "Provider" arises from the need to change the value in the trie, a process that modifies the trie's structure. Each instance of writing a non-default value (anything other than zero) to the state trie incurs higher gas costs due to the complexity of altering the trie's configuration. Hence, the act of registering a "Provider," which involves changing the state from 0 to 1, results in greater gas expenditure compared to registering a "Consumer," where the state remains unchanged at 0. In summary, the differential gas costs between registering as a "Provider" and a "Consumer" stem from the fundamental differences in state trie updates required for each role. The default value of 0 for "Consumer" roles typically avoids state changes, while the non-default value of 1 for "Provider" roles necessitates more gas-intensive state updates. \\

\begin{figure}[ht!]
    \centering
    \includegraphics[width=0.5\linewidth]{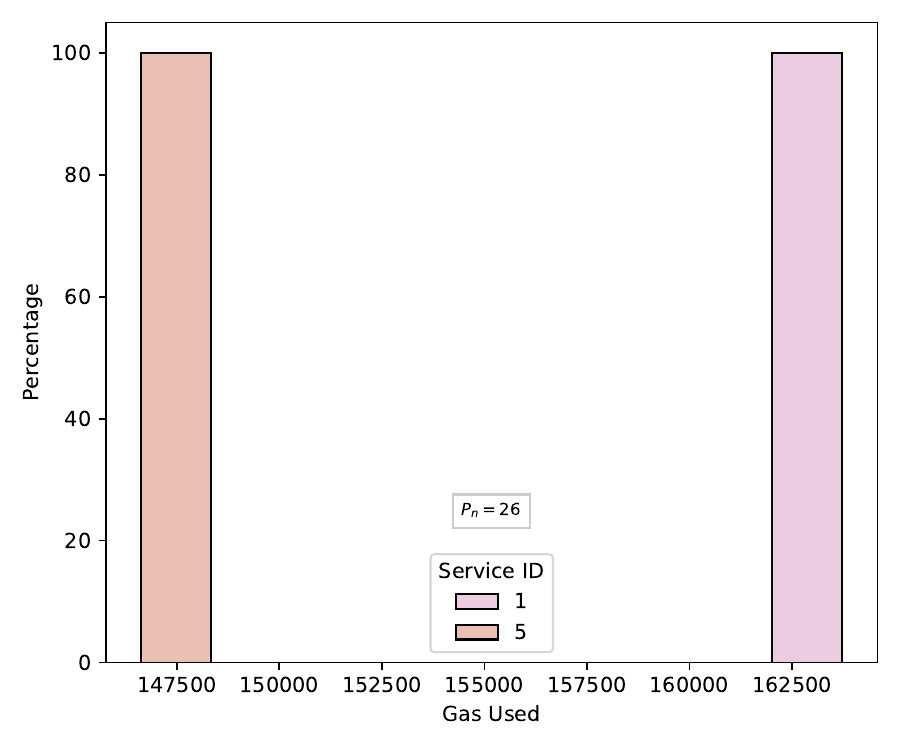}
    \caption{Gas Used for adding service batchsize $P_n$ = 26 across 10 iterations}
    \label{fig:gas_addservice}

    \vspace{1cm} 
    \captionof{table}{Gas Used and Percent for Service IDs 2, 3, 4, 5 for adding service batchsize $P_n$ = 26}
    \label{tab:gas-addservice}
    \begin{tabular}{p{1cm}p{1cm}p{1cm}}
        \toprule
        Service ID &  Gas Used &      Percent \\
        \midrule
                2 &    146629 &       100.00 \\ \midrule
                3 &    146629 &       100.00 \\ \midrule
                4 &    146629 &       100.00 \\
        \bottomrule
    \end{tabular}
\end{figure}

\subsubsection{Addition of a Service} 
In proposed inter-provider DApp, each provider can advertise up to five services. The \contract{AddService.sol} facilitates this by allowing providers to add details about each service they offer. This structure maps these services to their respective providers, ensuring each service can be uniquely identified and accessed. The analysis of gas usage for adding services, as depicted in Figure \ref{fig:gas_addservice}, focuses on the first and fifth services during their initial iteration when providers, denoted by $P_n$, are 26. The histogram provided elucidates the distribution of gas usage for two distinct service IDs, with each bar representing 100\% of observations for a respective service, situated across designated gas usage bins. 
For Service ID 1, the gas used demonstrates that all 26 service providers experience gas usage at the upper bin of approximately 162,500 units. This uniformity across providers at 100\% indicates a standardized requirement for gas consumption when initializing the first service. Such a high level of gas usage is attributed to the extensive system initialization required, including the comprehensive data mapping and storage operations essential for setting up the initial service environment\cite{wood2014ethereum}.
Conversely, Service ID 5 is characterized by a consolidation of all gas usage observations at a lower bin of around 147,500 units, also at 100\%. The histogram shows that initial service deployments require significantly more gas compared to subsequent additions. This difference in gas consumption results from the setup processes required for the first service. Specifically, setting up the service array and mappings for the first time involves accessing "cold" storage, which is more resource-intensive than the "warm" storage accesses that occur once the system is already initialized. Each transaction initiated from a provider using \function{addService} maps the service added to the specific service provider. The first entry into this mapping is particularly gas-intensive because it involves creating a new array within the mapping. Each new key in a mapping represents a transition from an uninitialized to an initialized state, necessitating high gas usage due to increased storage costs \cite{wood2014ethereum}. As additional services are added by the same provider, the gas costs is reduced since the mapping for that provider already exists after the first service is added, so subsequent additions involve fewer state changes compared to the initial setup: (i) accesses to already initialized entries in the array are "warm" and therefore less costly and (ii) subsequent entries do not require additional memory allocation or state initialization, reducing the gas costs further. This reduction in gas usage for services indexed as 2, 3, and 4 are evident in the overall distribution, as detailed in Table \ref{tab:gas-addservice}.

\begin{figure}
    \centering
    \includegraphics[width=0.7\textwidth]{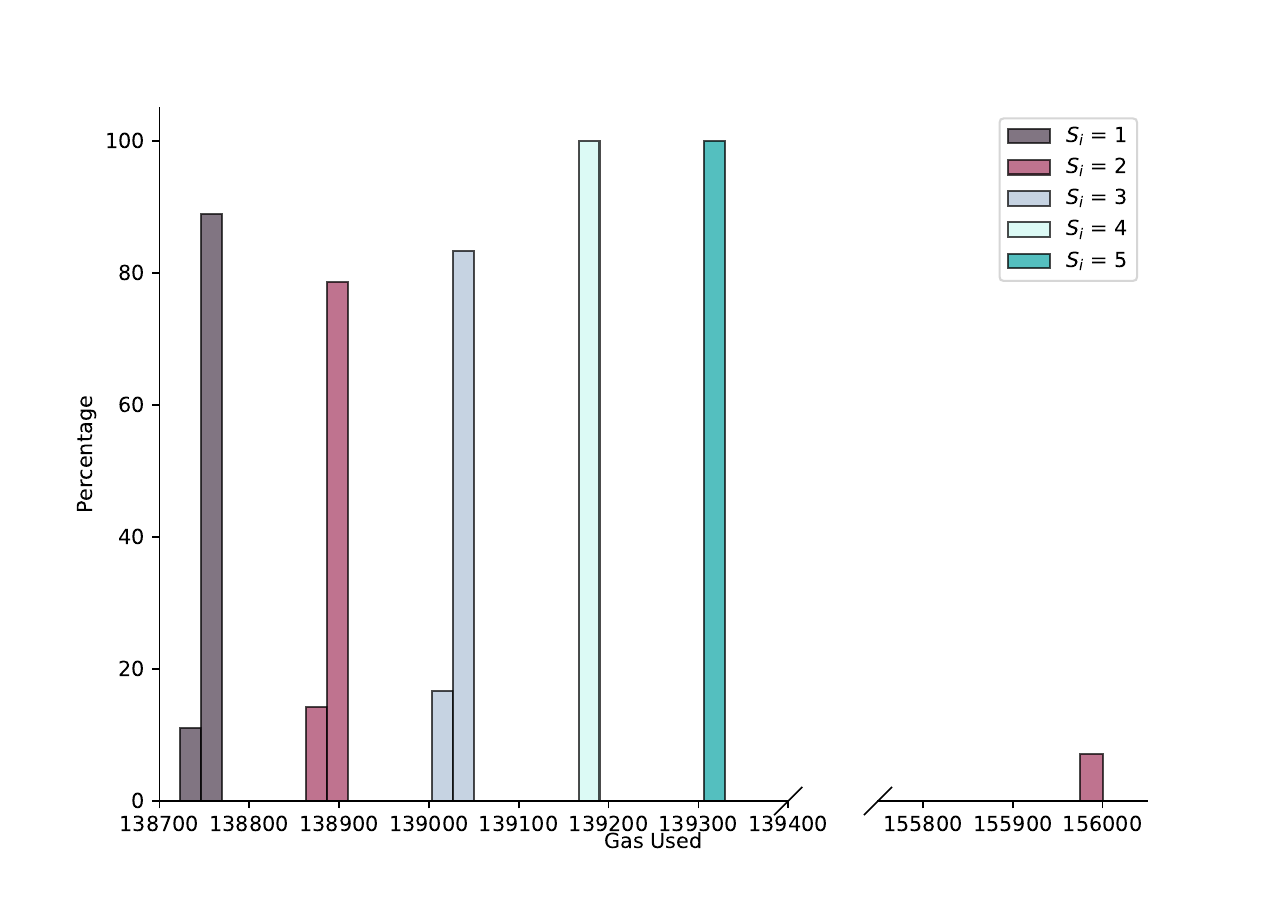}
    \caption{Gas Used for selection of a service when $P_n$ = 50 and $C_n$ = 50}
    \label{fig:gas-service-selection}
\end{figure}

\subsubsection{Selection of a Service}
\label{gas-explain-selection}
During the service selection phase in the proposed inter-provider DApp, each consumer can select one service from the available list of services up by all providers. The \contract{ServiceSelection} facilitates this by allowing providers to add details about each service they offer. 
As described in Table \ref{tab:smart_contracts}, involves two write functions and one read function within the \contract{ServiceSelection} smart contract. This process is crucial for the functioning of a DApp and involves both read and write operations that interact with the \contract{ServiceSelection} smart contract. Read operations are essential for allowing consumers to browse and evaluate the available services from different providers without altering the blockchain's state. These operations facilitate data retrieval, enabling users to make informed decisions. Conversely, write operations are necessary when a consumer decides to select a service. The service selection process begins with the consumer choosing a random provider from the available list. After selecting a provider, the consumer examines the specific services offered, ultimately making a selection. This is where write operations occur, as they involve recording the consumer's choices and mapping them to specific service IDs on the blockchain. The \function{getProviders} and \function{getServices} functions play a pivotal role in this process. The former retrieves a list of available providers, while the latter fetches the services provided by a selected provider. These functions ensure the smooth execution of the service selection process by enabling both data retrieval (read operations) and the recording of consumer selections (write operations).

When a service is selected for the first time, this event initializes a new entry in the smart contract selected services mapping. This initialization, known as cold access as explained earlier, is notably more gas-intensive compared to accessing already initialized entries (warm access). For example, in our scenario Service ID 2 is the first service selected by a consumer, it triggers a cold access. The observed gas usage for this initial selection was high at 155992 as shown in Figure \ref{fig:gas-service-selection}. Histogram in Figure \ref{fig:gas-service-selection}  presents a detailed analysis of the gas used across various service IDs, depicted in terms of percentages. This analysis provides insights into the distribution of gas usage for five distinct service IDs. It can be seen that Service ID 2 displays a more varied gas usage distribution, with three different values observed. Specifically, 138,880 units of gas account for 14.29\% of the instances, while 138,892 units represent the majority, with 78.57\% of the usage. Additionally, 155,992 units were used in 7.14\% of the instances. This high cost is because the process involves not only accessing the service from the blockchain but also initializing the mapping in the storage. This initialization is a critical factor and would similarly impact any service that is accessed for the first time.

However, looking at the Figure \ref{fig:gas-service-selection} it is important to highlight that the placement of a service within the provider’s list also influences gas costs. The further down the list a service is (e.g., Service ID 5), the more SLOAD operations are needed to reach and retrieve it. Each SLOAD operation has a cost, meaning that initialization plus retrieval can lead to progressively higher gas usage for services located deeper in the array. For example, Service ID 1, For Service ID 1, the gas usage is concentrated in two specific values. The gas used amounts to 138,740 units in 11.11\% of the instances where Service ID 1 was selected. In contrast, 138,752 units of gas were used in the remaining 88.89\% of instances for Service ID 1. Service ID 5 reports gas used values of 139312, reflecting the increased operational load due to its deeper position in the list. Service ID 4, slightly higher in the list, shows a gas usage of 139172. Service ID 3, even closer to the top, has gas used values of 139032 and 139020, showing a decrease in gas costs as the services are less deep in the list as it can be observed in Figure \ref{fig:gas-service-selection}. Service ID 3 shows a distribution between two values. The gas used amounts to 139,020 units in 16.67\% of the instances, while 139,032 units were used, accounting for 83.33\% of the instances. For Service ID 4, gas usage is uniformly 139,172 units in 100\% of instances, indicating no variability. Similarly, Service ID 5 shows a consistent gas usage of 139,312 units in 100\% of instances, suggesting no variation in gas consumption.

In the Ethereum, the cost of executing transactions can vary based on the composition of the data involved, particularly evident when service IDs are associated with different provider addresses within a DApp's service selection process. Variations in gas usage are observed when the same service ID is selected multiple times but linked to various provider addresses, showcasing minor discrepancies in gas costs
For instance, Service ID 1 has shown gas usage values of 138752 and 138740, Service ID 2 has shown values of 138880 and 138892, and Service ID 3 has values of 139032 and 139020. These variations are due to the differences in the number of zero and non-zero bytes in the provider addresses involved in the transactions.

Ethereum’s gas pricing guidelines specify that each zero byte of data or code in a transaction costs 4 gas units (\textit{Gtxdatazero}), whereas each non-zero byte costs 16 gas units (\textit{Gtxdatanonzero}) \cite{wood2014ethereum}.

Therefore, if a provider address, such as 7b980bb5399bf00043408d25dd96a8f1da389f34, contains more zero bytes compared to another address like ec208ff0cfc9dce2049bcc1e0d5994e53356dd98, transactions involving the former will be less expensive due to the lower number of costly non-zero bytes. This results in the observed fluctuations in gas costs—for example, a consistent decrease or increase by about 12 units—highlighting how the binary composition of provider addresses can influence the overall gas fees in blockchain operations. This underscores the impact of data structure on transaction costs in Ethereum, demonstrating that minor variations in byte composition can meaningfully affect gas expenses.

\begin{figure}
    \centering
    \includegraphics[width=0.5\textwidth]{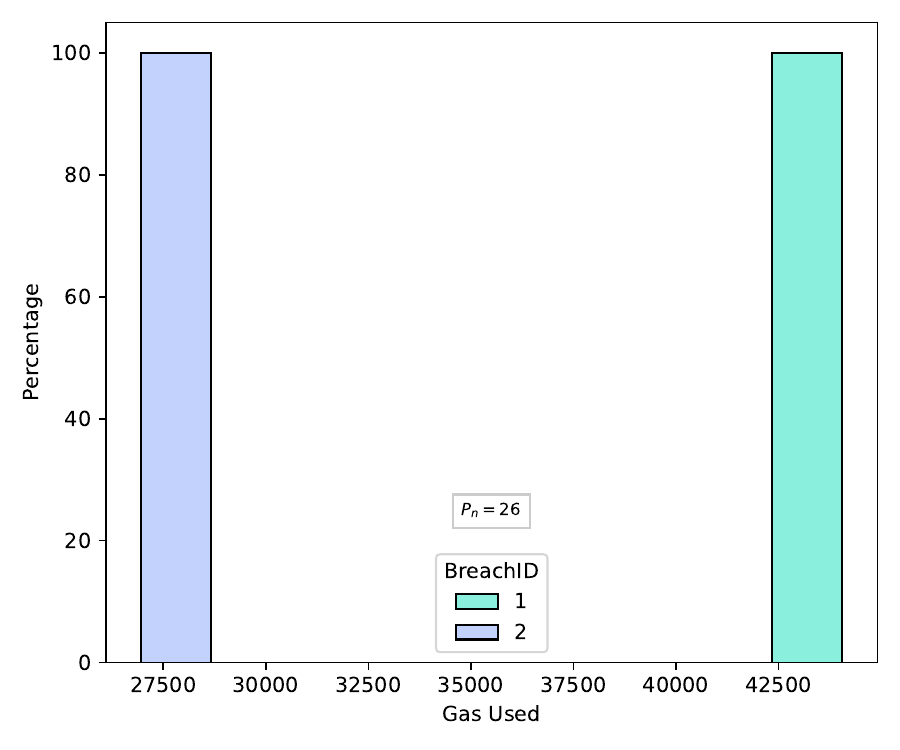}
    \caption{Gas Used for registering a breach $B_i$ of the ADs for batchsize 26 for providers}
    \label{fig:gas-breach}
\end{figure}

\subsubsection{Registration of Breach and Calculation of Penalty}

The \contract{RegisterBreach.sol} contract manages the registration of breaches and the \contract{CalculatePenalty.sol} calculation of penalties for service providers based on their number of breaches. It uses two mappings: \texttt{breaches}, which records the number of violations per provider, and \texttt{penalties}, which stores the corresponding financial penalties as explained in Section \ref{smart-contracts}.

The contract provides two main functions. The \function{registerBreach} function increments the breach count for a provider's account. It accepts \texttt{numBreaches}, a parameter that specifies the number of breaches to add, with a limit of \texttt{maxBreach} set to three. Once a provider reaches this limit, the \function{calculatePenalty} function within \contract{CalculatePenalty.sol} is automatically triggered, allowing penalty calculation to occur only when the breach cap is reached and reducing unnecessary gas usage. On the other hand, \textit{calculatePenalty} retrieves the current breach count from the \texttt{breaches} mapping. It then applies a computation: \(\texttt{penalty} = \texttt{fidelityFee} \times \texttt{breachCount}\).

Our analysis of gas consumption is visualized through a histogram in Figure \ref{fig:gas-breach} demonstrates the gas costs associated with the \function{registerBreach} function when breaches are introduced via smart contract used to fetch KPIs from Oracle. 

The gas cost for the initial breach registration is 44,058 units for each provider, reflecting the cost of writing to a new storage slot (cold access), as described earlier. Subsequent breaches registered for the same provider incur a reduced gas cost of 26,958 units because they access an existing storage slot (warm access). Figure~\ref{fig:gas-breach} shows two main bars. The left bar, at approximately 27,000 gas units, corresponds to subsequent registrations involving warm storage. The right bar, at about 44,000 gas units, corresponds to the initial cold storage access for the first breach registration.

When analyzing the gas consumption of the \function{calculatePenalty} function in the Solidity contract, which stands at 49,134 gas units, it's essential to consider the key activities that consume gas. The function then writes the computed penalty to the penalties mapping. Apart from that, Table \ref{table:gas_used_comparison} shows different gas used with base 12; this is because, as explained already above in Section \ref{gas-explain-selection}, Ethereum’s gas pricing guidelines specify that each zero byte of data or code in a transaction costs 4 gas units (\textit{Gtxdatazero}), whereas each non-zero byte costs 16 gas units (\textit{Gtxdatanonzero}) \cite{wood2014ethereum}. Since calculatePenalty function has associated provider addresses, therefore the provider with address as explained in \ref{gas-explain-selection}.

\begin{table}[t!]
\centering
\begin{threeparttable}
\caption{Gas Used for of Functions with Address Variations}
\label{table:gas_used_comparison}
\begin{tabular}{p{4cm}p{3cm}p{3cm}p{3cm}}
\toprule
\textbf{Function} & \textbf{Gas Used} & \textbf{Variance} & \textbf{BatchSize}  \\
\midrule
\function{calculatePenalty()} & 49134 & Base case & $P_n$= 50 \\
 & 46515 & 12 gas less\tnote{*}  & $P_n$= 5\\
\function{transferFunds()} & 31266 & Base case & $C_n$= 50 \\
 & 31254 & 12 gas less\tnote{*} & $P_n$= 5 \\
\bottomrule
\end{tabular}
\begin{tablenotes}
\item[*] Each 12 gas decrease corresponds to one fewer non-zero byte in the address parameter.
\end{tablenotes}
\end{threeparttable}
\end{table}

\subsubsection{Transfer of Funds}
In the contract \contract{TransferFunds.sol}, the \function{transferFunds} function is crucial for facilitating the secure transfer of Ether between ADs. 
Table \ref{table:gas_used_comparison} shows the gas used for the function. The gas consumption for transactions involving this function is influenced by the composition of the recipient's Ethereum address. As outlined in earlier above the transaction cost varies based on the data payload, specifically the zero and non-zero bytes in the address. Each non-zero byte in a transaction incurs a higher gas cost (16 units) compared to zero bytes (4 units), reflecting in varying total gas usage in transactions. For instance, an address with fewer zero bytes results in a gas usage of approximately 31,266 units, whereas an address with more zero bytes might only use about 31,254 units.

\begin{table*}[htbp]
\centering
\scriptsize
\caption{Statistical Summary of Observed Blockchain Parameters and Latency for \function{registerAD}}
\label{table-registeration-full-details}
\begin{adjustbox}{width=1\textwidth,center}
\begin{tabular}{c|c|cc|cc|cc|cc|cc}
\toprule
\textbf{Batch Size} & \textbf{Role} & \multicolumn{2}{c|}{\textbf{Gas Price (Gwei)}} & \multicolumn{2}{c|}{\textbf{Block Size (KB)}} & \multicolumn{2}{c|}{\textbf{Transaction Count}} & \multicolumn{2}{c|}{\textbf{Mempool Time (s)}} & \multicolumn{2}{c}{\textbf{Latency (s)}} \\
\cmidrule(r){3-4} \cmidrule(r){5-6} \cmidrule(r){7-8} \cmidrule(r){9-10} \cmidrule(r){11-12}
 &  & \textbf{Mean} & \textbf{Std} & \textbf{Mean} & \textbf{Std} & \textbf{Mean} & \textbf{Std} & \textbf{Mean} & \textbf{Std} & \textbf{Mean} & \textbf{Std} \\
\midrule
2 & Consumer & 15.99 & 4.43 & 116.10 & 30.28 & 129.20 & 25.21 & 12.14 & 4.14 & 13.68 & 4.05 \\ 
2 & Provider & 15.99 & 4.43 & 116.10 & 30.28 & 129.20 & 25.21 & 12.16 & 4.17 & 13.73 & 4.10 \\ \midrule
16 & Consumer & 20.00 & 1.21 & 92.74 & 18.88 & 132.59 & 23.36 & 10.03 & 2.27 & 15.92 & 3.29 \\ 
16 & Provider & 19.97 & 1.22 & 95.04 & 19.89 & 136.40 & 27.37 & 10.86 & 1.95 & 15.82 & 2.76 \\ \midrule
30 & Consumer & 44.21 & 7.29 & 99.11 & 39.13 & 156.95 & 42.60 & 10.20 & 3.77 & 21.02 & 6.15 \\
30 & Provider & 44.29 & 7.22 & 100.10 & 41.02 & 160.00 & 43.05 & 11.05 & 3.24 & 20.67 & 6.61 \\ \midrule
44 & Consumer & 75.24 & 11.42 & 124.12 & 55.22 & 176.48 & 50.23 & 12.57 & 5.89 & 25.92 & 8.98 \\
44 & Provider & 74.89 & 11.59 & 127.81 & 60.13 & 179.30 & 52.18 & 13.00 & 4.89 & 25.68 & 9.10 \\ \midrule
58 & Consumer & 4.59 & 0.48 & 166.67 & 75.91 & 200.20 & 54.30 & 13.01 & 4.96 & 28.35 & 8.61 \\
58 & Provider & 4.59 & 0.50 & 176.59 & 79.48 & 203.45 & 56.75 & 12.85 & 5.41 & 29.24 & 9.31 \\ \midrule
72 & Consumer & 12.06 & 4.87 & 160.16 & 62.11 & 215.67 & 60.12 & 13.93 & 5.16 & 32.27 & 9.96 \\
72 & Provider & 12.03 & 4.88 & 169.75 & 66.10 & 218.33 & 61.40 & 14.90 & 5.23 & 33.35 & 9.19 \\ \midrule
86 & Consumer & 19.01 & 3.55 & 136.46 & 60.13 & 225.12 & 65.30 & 12.99 & 5.44 & 32.15 & 11.39 \\
86 & Provider & 18.97 & 3.55 & 129.74 & 61.54 & 230.40 & 66.50 & 13.21 & 5.24 & 34.18 & 11.69 \\ \midrule
100 & Consumer & 20.65 & 4.17 & 127.24 & 65.45 & 240.11 & 70.25 & 14.36 & 9.64 & 36.32 & 15.18 \\
100 & Provider & 20.65 & 4.10 & 123.94 & 64.15 & 245.32 & 72.18 & 13.64 & 7.08 & 37.84 & 13.77 \\
\bottomrule
\end{tabular}
\end{adjustbox}
\end{table*}

\subsection{Concurrency-Driven Execution Time}
\label{sec:concurrency-latency}

\begin{figure}
    \centering
    \includegraphics[width=0.7\textwidth]{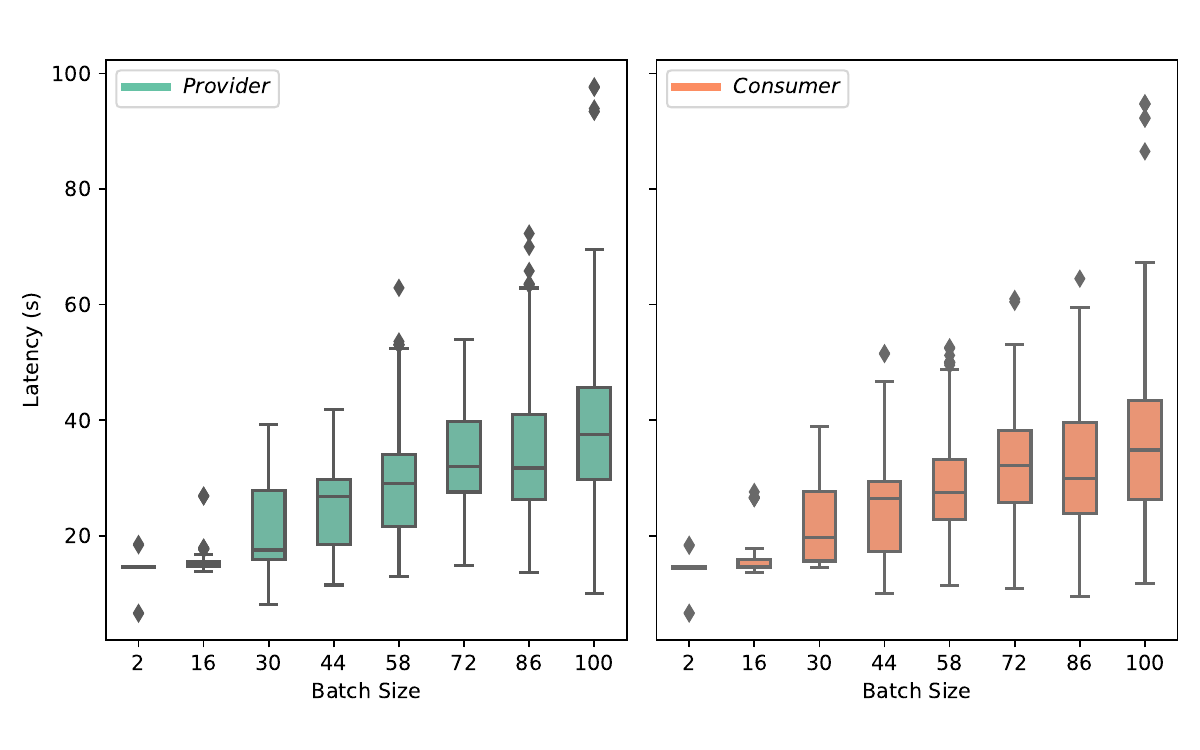}
    \caption{Latency for registration of the ADs for batchsize increasing from 2 to 100 with step 14}
    \label{fig:Latency-box-registeration}
\end{figure}

\subsubsection{Registration of an AD}

We conducted a series of tests on the Sepolia testnet to evaluate the performance of the registration function explained earlier and mentioned in Table \ref{tab:smart_contracts} within our DApp, focusing specifically on different batch sizes and different roles assigned. These tests simulated the registration process for batches ranging from 2 to 100, with increments of 14, where each batch size was evenly split between providers and consumers. For example, a batch of 2 consisted of one provider and one consumer, while a batch of 16 comprised eight providers and eight consumers. Each scenario was repeated ten times to ensure statistical significance.

Figure \ref{fig:Latency-box-registeration} presents box plots that compare the total latency (as defined in Section \ref{define-latency}) for both providers and consumers across various batch sizes. Here, “total latency” encompasses both the time spent in the mempool waiting to be included in a block and the on-chain execution time once the transaction is mined. Analyzing latency across different batch sizes, we find that providers (\function{registerAD} = 1) on average experience slightly higher latency than consumers (\function{registerAD} = 0)—for instance, 29.40 seconds for consumers and 31.50 seconds for providers overall. At small batch sizes like 2, their latencies remain close (13.68s vs. 13.73s), but at batch size 100, providers reach 37.84s compared to 36.32s for consumers. This difference arises from provider registrations requiring more intensive state changes, converting zero-value storage to non-zero \cite{wood2014ethereum}.

The results are further supported by Table \ref{table-registeration-full-details}, which details various metrics (e.g., gas price, block size, transaction count, mempool time). 
We also observed certain batches (e.g., 16, 30, 44) where providers’ transactions exhibit lower latencies than consumers, despite conventional expectations. This reversal arises from variations in gas price per unit within those batches, making the higher-gas-consuming provider transactions more attractive to validators seeking increased block rewards. As batch sizes increase beyond 44, the cumulative effects of blockchain network congestion become more pronounced \cite{gencer2018decentralization}. Consequently, the computational overhead in provider transactions—requiring additional state updates—further amplifies latency at high concurrency levels (e.g., from 58 up to 100).

\begin{figure}
    \centering
    \includegraphics[width=0.7\textwidth]{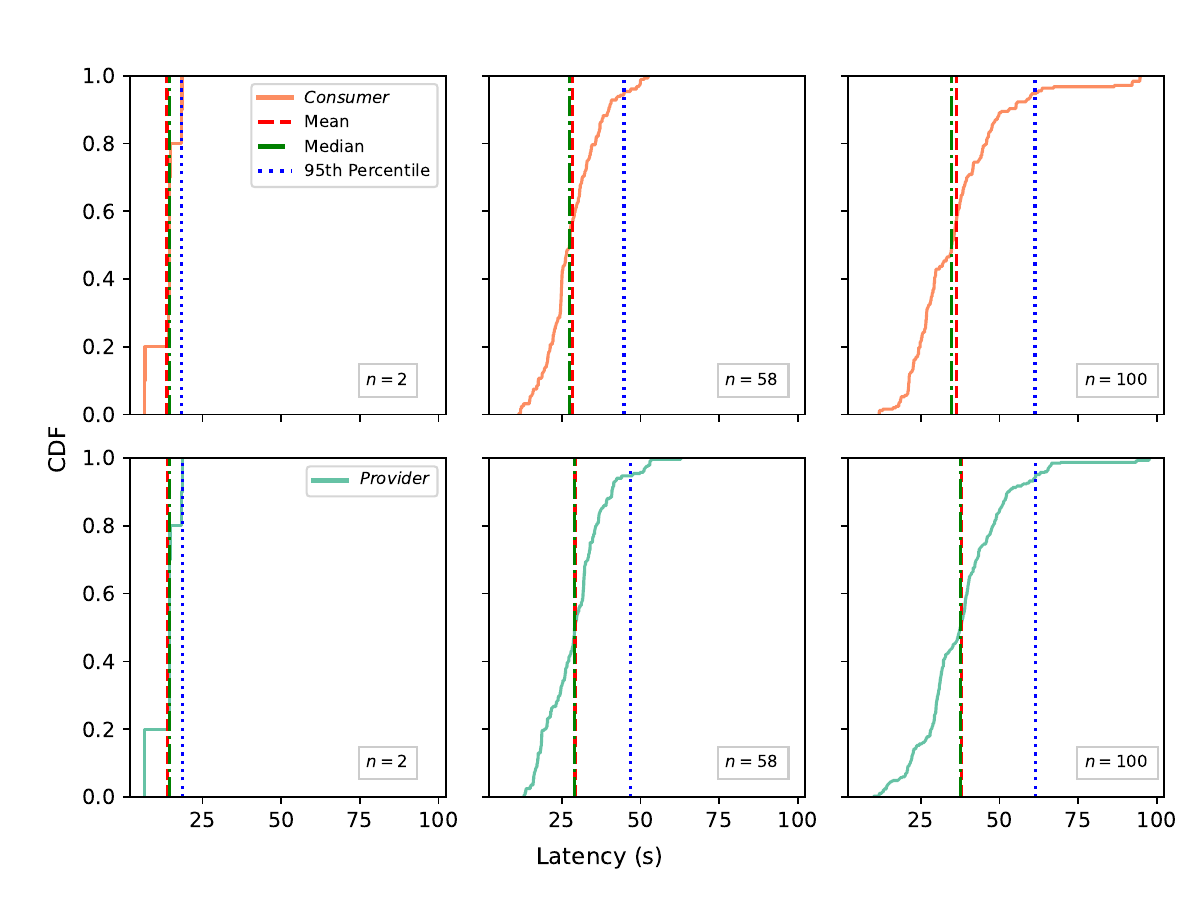}
    \caption{Latency for registration of the ADs with batchsize $n$ = 2, 58 and 100}
    \label{fig:Latency-cdf-registeration}
\end{figure}

\begin{figure}
    \centering
    \includegraphics[width=0.5\textwidth]{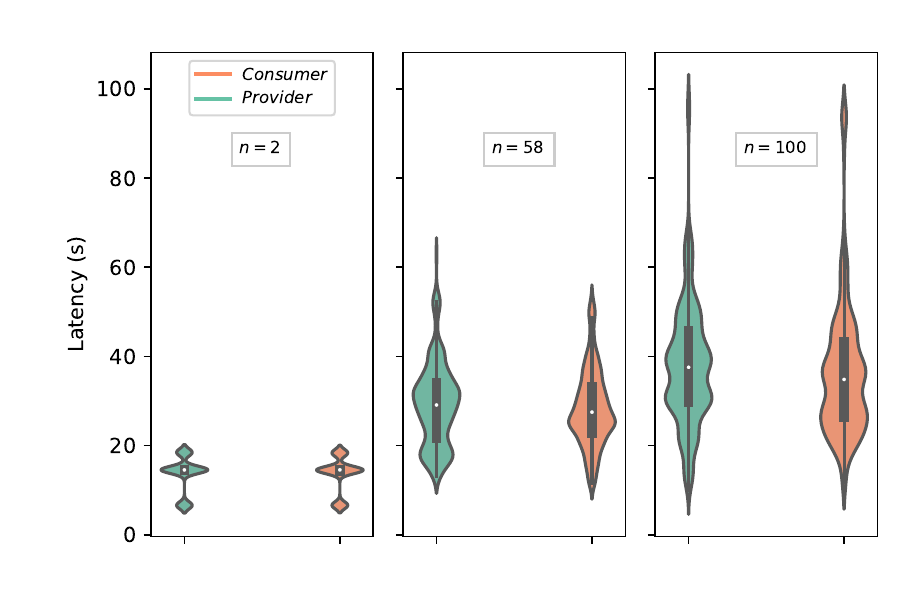}
    \caption{Violin plot for registration of the ADs with batchsize $n$ = 2, 58 and 100}
    \label{fig:violin_registeration}
\end{figure}

Figures \ref{fig:Latency-cdf-registeration} and \ref{fig:violin_registeration} offer complementary insights on how latency distributions evolve. The CDF plots highlight the probabilistic distribution of latencies for batch sizes of 2, 58, and 100, identifying mean, median, and 95th percentile markers. As batch size increases, the latency distribution broadens and shifts to higher values, signaling reduced performance. Similarly, the violin plots capture both central tendencies and the extent of outliers, emphasizing the variability at larger batch sizes. This resonates with the box-plot findings: even if mempool time remains in a moderate range (around 10–15 seconds for most transactions), total registration latency rises significantly due to on-chain execution constraints once blocks fill with concurrent transactions.

Both the CDF and violin plots offer complementary insights, with the violin plots providing a visual emphasis on the variability and wider distributions observed in larger batches. This dual approach highlights the critical influence of batch size on the registration times of ADs, reflecting both the general and detailed aspects of latency behaviors across the network.

\begin{figure}
    \centering
    \includegraphics[width=0.5\textwidth]{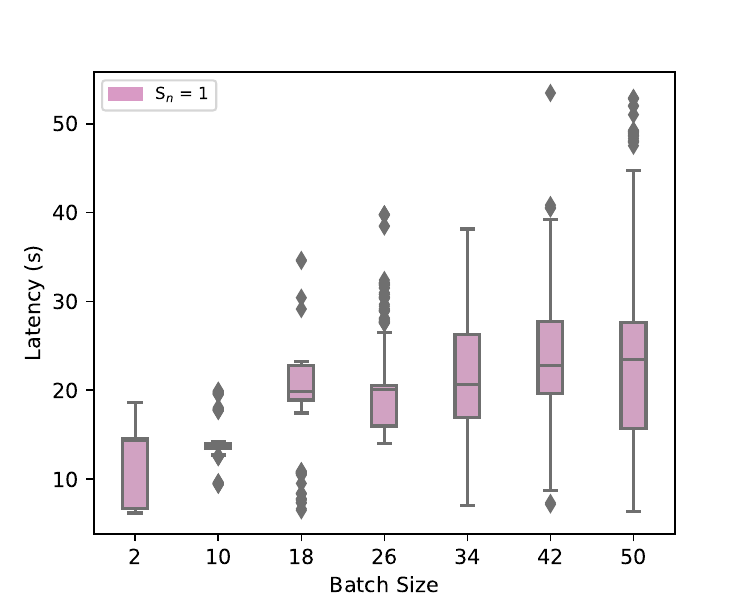}
    \caption{Box plot for adding service of the Provider AD with batchsize $n$ = 2 to 50}
    \label{fig:box-1addService-2-50x10}
\end{figure}
\subsubsection{Addition of a Service} 
To evaluate the performance of the service addition function described earlier and referenced in Table \ref{tab:smart_contracts} within our DApp, we conducted a series of tests on the Sepolia testnet, focusing specifically on various batch sizes for the addition of a single service. These tests simulated the registration process with batch sizes ranging from 2 to 50, increasing in increments of 8. \\
Additionally, to understand the impact of adding up to five services, we performed simulations with batch sizes of 2, 26, and 50. Due to the high cost associated with multiple service additions in large batches and multiple repetition to generate results with statistical significance, we limit this analysis to only three batches to economize on resources while still gaining insights into latency behaviors under heavier loads.

\begin{table}[tp!]
\centering
\caption{Statistical Summary of Observed Blockchain Parameters and Latency for \function{addService}}
\begin{tabular}{p{4cm}p{2cm}p{2cm}p{2cm}p{2cm}}
\toprule
\textbf{Metric} & \textbf{Mean} & \textbf{25th\%} & \textbf{Median} & \textbf{95th\%} \\
\midrule
\textbf{Gas Price (Gwei)} & 9.89 & 7.29 & 8.86 & 15.42 \\
\textbf{Block Size (KB)} & 166.92 & 106.11 & 152.58 & 312.38 \\
\textbf{Transaction Count} & 134.09 & 114.00 & 123.00 & 202.00 \\
\textbf{Latency (s)} & 21.68 & 16.45 & 20.61 & 35.40 \\
\bottomrule
\end{tabular}

\label{table:addservice-1}
\end{table}

Figure \ref{fig:box-1addService-2-50x10} presents box plots illustrating the latency of adding one service as shown as $S = 1$ in batches 2 to 50 with step 8. Figure \ref{fig:box-1addService-2-50x10} reveals an overall increasing trend in latency as the batch size increases from 2 to 50. Additionally, Table \ref{table:addservice-1} presents a comprehensive summary of quantitative data categorized by various batch sizes.

Upon observation the Figure \ref{fig:box-1addService-2-50x10} and Table \ref{table:addservice-1}, it can be noticed that batch sizes 10, 18, and 26 reveals the lowest Interquartile Range (IQR) indicating minimal variability despite the same operational function being performed.

Examining gas prices for these transactions it becomes evident that batch sizes 10, 18, and 26 exhibit moderate to low variability in gas prices, with standard deviations of 1.87 Gwei, 0.60 Gwei, and 1.28 Gwei, respectively. This stability in gas prices is crucial for minimizing unpredictability in transaction prioritization \cite{buterin2014ethereum,javed2025empiricalsmartcontractslatency}. When gas prices remain consistent, validators are less inclined to reorder transactions based on fluctuating fee incentives, leading to more predictable processing times and reduced latency variability \cite{buterin2019eip1559}. Such stability prevents sudden spikes in transaction fees from disrupting processing times, thereby contributing to the observed consistency in latency.

Block size, another important factor influencing network performance, varies across batch sizes. Notably, Batch Size 10 has the largest mean block size of 217,916 bytes with a standard deviation of 72,016 bytes, whereas Batches 18 and 26 maintain smaller block sizes of 152,213 bytes and 156,113 bytes, with standard deviations of 46,284 bytes and 59,352 bytes, respectively. While one might initially assume that larger block sizes would lead to higher latency due to increased propagation and validation times, the data indicates otherwise. Despite Batch Size 10's larger mean block size, there is no corresponding increase in latency variability. This suggests that factors beyond block size variability, such as the network's capacity to handle larger blocks or the specific nature of the transactions being processed, contribute to the stable latency observed \cite{croman2016scaling,javed2025empiricalsmartcontractslatency}.

Transaction count further influences latency dynamics. Batch Size 10 records the lowest mean transaction count of 115.3 with a standard deviation of 9.85, while Batches 18 and 26 handle higher transaction volumes of 142.63 and 138.64 transactions, respectively, with moderate variability (standard deviations of 32.00 and 24.72). Lower and more predictable transaction counts, as observed in Batch Size 10, reduce the processing burden on the network, leading to more stable latency \cite{gervais2016security}. However, Batches 18 and 26 manage higher transaction volumes without significantly increasing latency, indicating that the Sepolia testnet can handle larger transaction volumes through effective scaling mechanisms and optimized consensus algorithms \cite{kwon2018scalability}.  

Comparatively, other batch sizes such as 2, 34, 42, and 50 exhibit higher variability in latency, which can be attributed to less optimal balances between gas price stability, block size, and transaction count. For instance, Batch Size 2, despite having a relatively low transaction count and moderate gas price variability, shows higher latency variability than Batch Size 10, which could be attributed to network congestion. Larger batch sizes like 42 and 50, with greater gas price and transaction count variability, experience significantly higher latency standard deviations. It's important to note that higher gas prices typically deter excessive transaction submissions, potentially reducing congestion. However, this effect can be offset if the increased transaction counts at higher batch sizes overwhelm the network capacity. This suggests that network congestion is not solely dependent on gas prices but also significantly influenced by the transaction volume and network throughput capacity \cite{eyal2014majority, decker2013information}.

Lastly, Batch Size 42 exhibits slightly higher average latency than Batch Size 50. Batch Size 42 consistently has higher gas prices, causing greater competition and potential congestion, which further elevates latency. In contrast, Batch Size 50, with a decreased mean gas price from 12 Gwei to 8 Gwei, experiences less congestion resulting in more predictable and stable network performance. 

\begin{figure}[htp]
    \centering
    \includegraphics[width=0.5\linewidth]{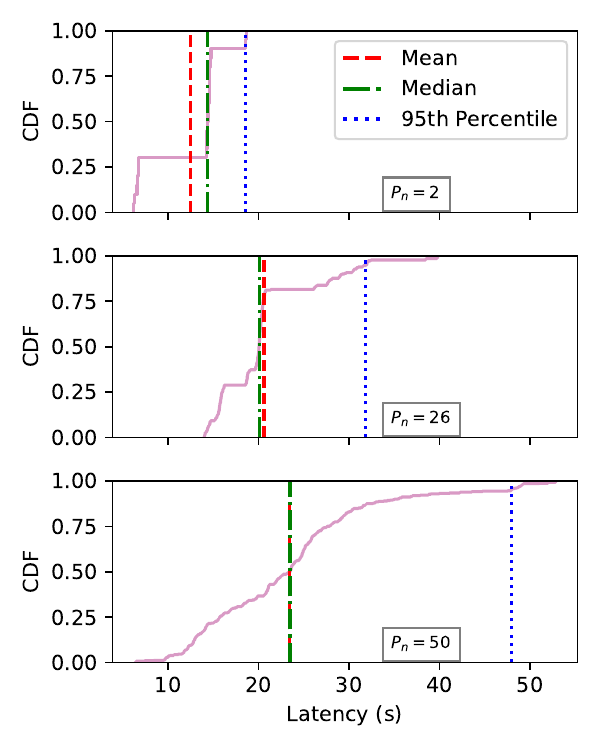}
    \caption{CDF for adding service of the Provider AD with batchsize $n$ = 2 to 50}
    \label{fig:CDF-1addService-2-50x10}
\end{figure}

\begin{table}[h]
\centering
\caption{Statistical Summary of Transaction Latencies by Service ID}
\label{tab:service-ID-latency}
\begin{tabular}{cccc}
\toprule
\textbf{Service ID} & \textbf{Mean (s)} & \textbf{Median (s)} & \textbf{95th Percentile (s)} \\
\midrule
1          & 22.81    & 21.20      & 35.68               \\
2          & 22.86    & 22.10      & 34.65               \\
3          & 21.84    & 21.00      & 35.33               \\
4          & 21.63    & 20.96      & 33.92               \\
5          & 21.94    & 20.97      & 35.43               \\
\bottomrule
\end{tabular}
\end{table}

\begin{figure}[tp!]
    \centering
    \subfloat[Boxplot for service selection function with batch size ranging from $C_n$ = 2 to 50]{%
        \includegraphics[width=0.48\linewidth]{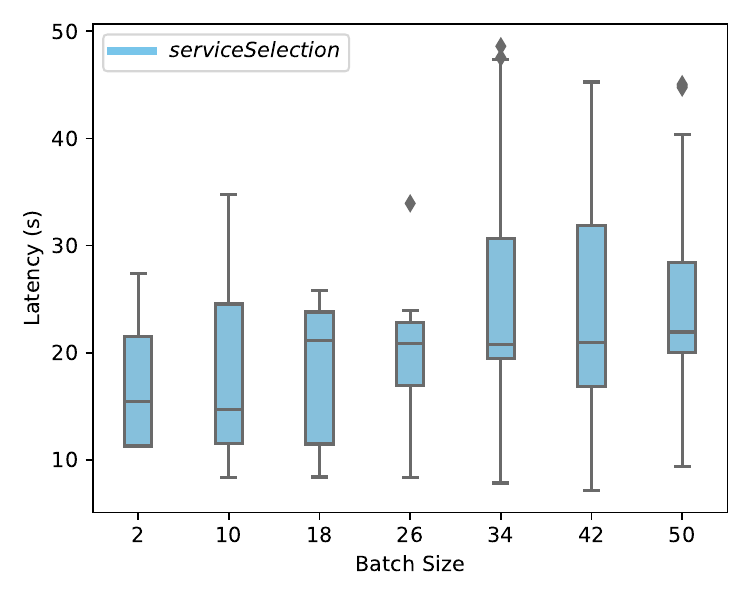}%
        \label{fig:boxplot-write-selection}%
    }
    \hfill
    \subfloat[Violin plot for selected service ID with batch size ranging from $C_n$ = 2 to 50]{%
        \includegraphics[width=0.48\linewidth]{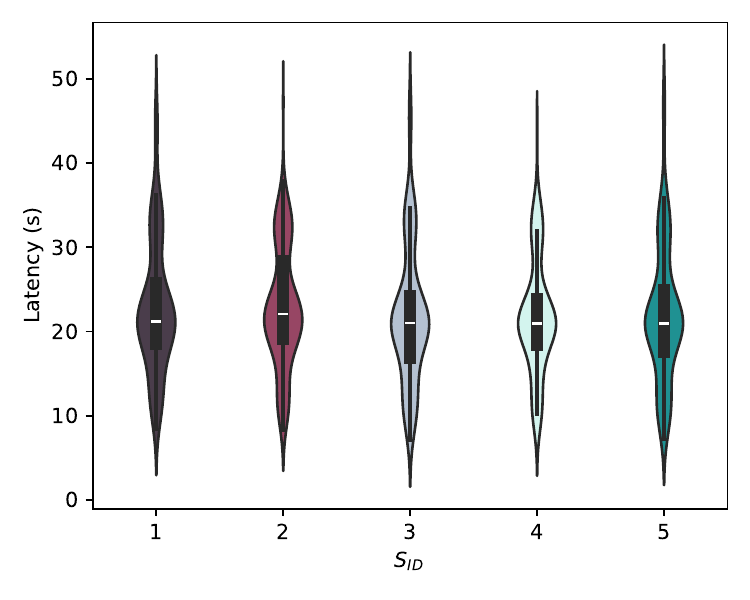}%
        \label{fig:boxplot-service-ID}%
    }
    \caption{Comparison of Service Selection Latency and Service ID Latency across batch sizes.}
    \label{fig:combined-service-plots}
\end{figure}

\begin{table}[tp!]
\centering
\caption{Statistical Summary of Observed Blockchain Parameters and Latency for \function{serviceSelection}}
\begin{tabular}{p{5cm}p{2cm}p{2cm}p{2cm}p{2cm}}
\toprule
\textbf{Metric} & \textbf{Mean} & \textbf{25th\%} & \textbf{Median} & \textbf{95th\%} \\
\midrule
\textbf{Gas Price (Gwei)} & 22.98 & 20.34 & 22.43 & 30.56 \\
\textbf{Block Size (KB)} & 182.19 & 92.57 & 116.36 & 633.55 \\
\textbf{Transaction Count} & 123.17 & 105.00 & 115.00 & 193.00 \\
\textbf{Latency (s)} & 22.21 & 18.40 & 21.13 & 35.30 \\
\bottomrule
\end{tabular}
\label{table:selectservice}
\end{table}

\subsubsection{Selection of a Service}
The smart contract managing the service selection as \contract{SelectService.sol} includes write functions as explained in Table \ref{tab:smart_contracts}.

The latency for \function{serviceSelection} function in smart contract \contract{SelectService.sol} across different batch sizes is visualized in Figure~\ref{fig:boxplot-write-selection}, presenting a boxplot for batch sizes ranging from 2 to 50. Service selection latency refers to the time elapsed from the initiation of a service selection by a consumer—such as selecting service ID $i$ from a provider—until the transaction is recorded on the blockchain.

The Figure~\ref{fig:boxplot-write-selection} as well as the data presented in Table~\ref{table:selectservice} illustrate an increase in mean latency (s) as the batch size increases, ranging from 17.40s to 23.38s.

In Section~\ref{fig:gas-service-selection}, we discussed that adding a service to each provider's service array results in varying amounts of gas used, depending on the position of that service ID, which ranges from 1 to 5.

However, in terms of the selection process, the latencies for each service, as indicated by the median latency, do not show significant differences as depicted in Figure~\ref{fig:boxplot-service-ID}, as well as Table \ref{tab:service-ID-latency} where each $S_{ID}$ and their selection in all transactions are shown as well as mean, median and 95\% percentile. 

We observe the transaction data it is observed that the transaction count peaks at batch size 2, while for the rest of batch sizes, it shows a reduced range of variation as batch size increases. However, larger batch sizes, particularly 34 and 42, exhibit a wider range of latencies. For example, a consumer selecting Service ID 5 from a provider in batch 34 experienced a latency exceeding 30 seconds at a gas price of 17.43 Gwei, despite being included in the same block.

Another critical metric is the gas price, measured in Gwei. Throughout the ten repetitions for each batch size the gas price remained stable, ranging from 20.5 to 23.26. This suggests that validators did not need to prioritize transactions based on gas prices \cite{wood2014ethereum}, resulting in relatively consistent latencies across different batches. This consistency is reflected in median latencies presented Figure \ref{fig:boxplot-service-ID} and Figure~\ref{fig:boxplot-write-selection}. However, while batch size increases lead to greater variability, the median latency observed does not show significant fluctuations.

\subsubsection{Registration of Breach and Calculation of Penalty}
The \function{registerBreach} function in the \contract{RegisterBreach.sol} smart contract plays a crucial role in the enforcement of SLAs by incrementing the breach count for a provider. Specifically, when invoked, it increases the number of breaches recorded for the associated provider's address by the specified \texttt{numBreaches}. The contract sets a maximum breach limit, denoted as \( B_{\text{max}} \), with a threshold value of 3. This limit allows the contract to effectively track and manage the cumulative breaches for each provider. Once a provider's breach count reaches this maximum, the \function{calculatePenalty} function is automatically triggered to assess and apply penalties. This system ensures that all breaches are accounted for and penalized accordingly, upholding the terms and integrity of the SLAs.

\begin{figure}
 \centering
    \includegraphics[width=0.5\linewidth]{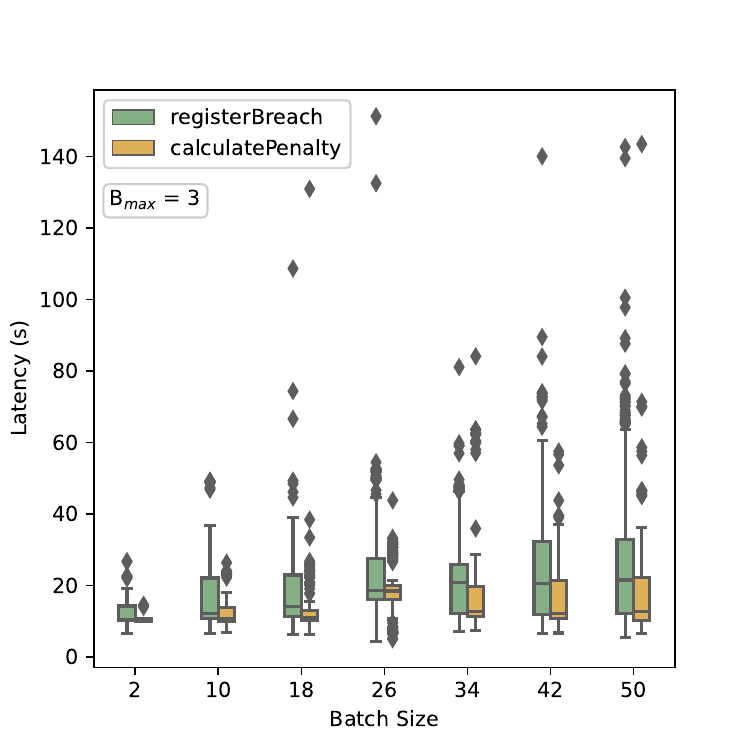}
    \caption{Boxplot for registerBreach $P_n$ = 2 to 50}
    \label{fig:boxplot-write-breach-max}
\end{figure}

Figure \ref{fig:boxplot-write-breach-max} illustrates the registration of all breaches, where the maximum breaches $B_{max}$ = 3), across various provider batch sizes, are shown increasing. Additionally, the triggered \function{calculatePenalty} function, which computes the total amount of penalty to be paid.

In the same figure, we observe a clear depiction of how latency in blockchain transaction processing varies with different batch sizes. Providers register a total of three batches using the \function{registerBreach} function, with batch sizes ranging from 2 to 50. Another boxplot within the figure displays the \function{calculatePenalty} function, triggered once three breaches have been recorded. At a batch size of 2, the median latency, represented by the line within the box, is around 10 seconds, which suggests a balanced performance with half of the latencies below this value and half above. The box, showing the IQR from approximately 8 to 12 seconds, indicates that the middle 50\% of data points are tightly clustered around the median, demonstrating consistent transaction processing times. Outliers, depicted as diamonds, occur at around 15 seconds.

However, as the batch size increases to 10, the median latency rises slightly to about 12 seconds, reflecting an increase in processing time due to increased transactions. The IQR expands to between 10 and 16 seconds, showing greater variability in how long transactions take to complete. This spread suggests that as more transactions are processed simultaneously, the likelihood of experiencing variable processing times increases.

Further increases in batch size continue this trend. For example, at a batch size of 18, the median latency extends to 14 seconds, and the IQR widens further, indicating that the range of typical transaction times is broadening as the system handles more data. The presence of more outliers. By the time we reach larger batch sizes of 26 to 50, we can observe that while the median latencies increase gradually—indicating a general rise in the average time transactions take—the outliers reaching as high as 140 seconds reveal that under certain conditions, some transactions take an exceptionally long time to process illustrate a network that is straining under the load.

To understand this, we also have a look at the other parameters such as block size (KB), transaction count as well as gas price. The function \function{registerBreach} represents three separate transactions or breaches, consistently shows higher transaction counts and block sizes compared to \function{calculatePenalty}, which is triggered as a result of these breaches but only represents one transaction when a maximum of three breaches are recorded. This pattern holds across all batch sizes, with \function{registerBreach} not only having higher average transaction counts but also exhibiting more variability in these numbers. Similarly, the block sizes for \function{registerBreach} are notably larger.

As batch sizes increased from 10 onwards, both \function{registerBreach} and \function{calculatePenalty} saw increase in transaction counts and block sizes, with \function{registerBreach} consistently higher in both metrics. Gas prices surged notably when batch sizes hit 18, peaking above 50 Gwei, which is related with network load at that time \cite{werner2020step} by batch size 42. Similarly, latency for \function{registerBreach} increased with larger batch sizes, demonstrating its higher demand on network resources. For instance, at batch size 42, \function{registerBreach} recorded a block size of 230.53 KB and 131.61 transactions, higher than \function{calculatePenalty}, which showed a block size of 239.88 KB and 119.55 transactions. This pattern highlights the increased load \function{registerBreach} places on the system, reflecting its complex transactional requirements compared to \function{calculatePenalty}.

\begin{table}[tp!]
\centering
\caption{Statistical Summary of Observed Blockchain Parameters and Latency for Key Functions.}
\begin{threeparttable}
\adjustbox{max width=\textwidth}{
\begin{tabular}{l*{12}{c}}
\toprule
\multirow{2}{*}{\textbf{Metric}} & \multicolumn{4}{c}{\function{registerBreach}} & \multicolumn{4}{c}{\function{calculatePenalty}} & \multicolumn{4}{c}{\function{transferFunds}} \\
\cmidrule(lr){2-5} \cmidrule(lr){6-9} \cmidrule(lr){10-13}
 & \textbf{Mean} & \textbf{25th\%} & \textbf{Median} & \textbf{95th\%} & \textbf{Mean} & \textbf{25th\%} & \textbf{Median} & \textbf{95th\% }& \textbf{Mean} & \textbf{25th\%} & \textbf{Median} & \textbf{95th\%} \\
\midrule
\textbf{Gas Price (Gwei) }   & 30.53 & 14.77 & 24.37 & 79.66 & 29.03 & 15.23 & 23.72 & 84.38 & 5.98  & 5.50  & 5.85  & 7.88  \\
\textbf{Block Size (KB)}     & 237.47 & 124.86 & 185.07 & 703.18 & 234.71 & 122.45 & 182.71 & 756.87 & 161.93 & 102.49 & 141.03 & 313.31 \\
\textbf{Transaction Count }  & 125.07 & 100.00 & 115.00 & 198.00 & 120.20 & 96.00  & 110.00 & 191.00 & 160.76 & 130.00 & 151.00 & 276.00 \\
\textbf{Latency (s)}         & 21.88 & 11.91  & 19.63  & 45.96  & 16.20  & 10.61  & 12.68  & 30.61  & 28.18  & 18.03  & 27.63  & 53.34  \\
\bottomrule
\end{tabular}
}
\label{table:combined_summary}
\end{threeparttable}
\end{table}

\begin{figure}[htb]
    \centering
    \includegraphics[width=0.5\linewidth]{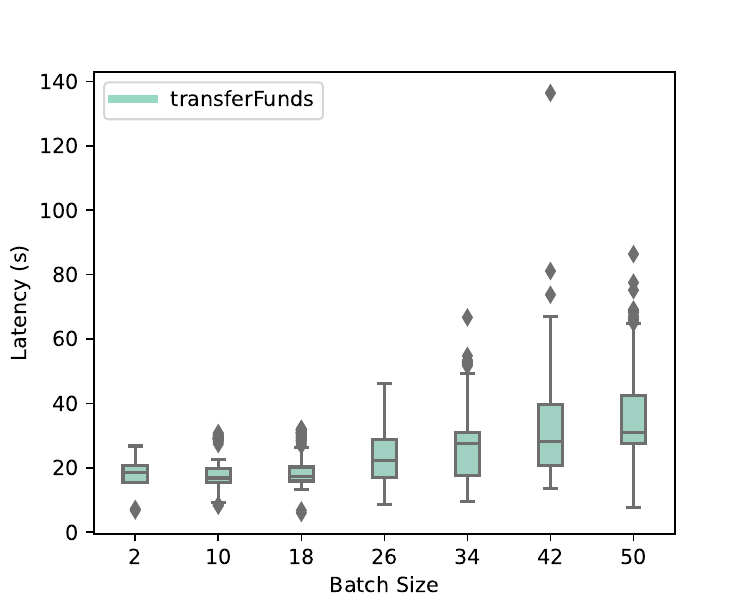}
    \caption{Boxplot for all functions after breach}
    \label{fig:boxplot-transfer}
\end{figure}


\subsubsection{Transfer for Funds}
For the final phase, the amount calculated can be transferred to provider's wallet using \function{tranferFunds}. This function takes as input the provider's address and sends a specified amount of Ether to that address. The amount of Ether sent is specified in the script.  To complete, we analyze, the batch size when funds are transferred from each consumer address to a provider address. 
The analysis of median latencies across varying batch sizes is provided in Figure \ref{fig:boxplot-transfer}. 

Starting with a median latency of 16.54 seconds for batch size 2, latency increases to 18.86 seconds by batch size 10. This upward trend continues, slightly rising to 17.39 seconds for batch size 18 and jumping to 22.27 seconds by batch size 26. This pattern marks the onset of a performance bottleneck as larger transaction volumes begin to adversely affect latency. As batch sizes further expand to 34, 42, and 50, median latency escalates faster, reaching a peak of 31.03 seconds. 

We observed the following while examining the blockchain network parameters for transactions across all batch sizes collectively: The range of gas prices spans from approximately 2.79 to 8.42 Gwei, with an average gas price of around 5.98 Gwei. The standard deviation of the gas prices is 0.92 Gwei, indicating a moderate spread around the mean. The block sizes vary significantly, ranging from about 55.14 KB to 445.70 KB. The average block size is approximately 161.93 KB, and the standard deviation is 77.02 KB, showing considerable variation in block size across different transactions. Furthermore, the number of transactions per block ranges from 65 to 318, with an average of approximately 160.76 transactions per block. The standard deviation here is 47.08, suggesting a diverse range of transaction counts per block. These statistics provide insights into the typical transaction and block characteristics within the dataset, highlighting that the observed higher average mean latency (s) can be attributed to factors such as lower gas prices (Gwei) as well as high mean transactions per block and block size.

Lastly, \function{transferFunds} exhibits a steady climb in median latency as batch sizes grow, from 16.54 s at batch size 2 to over 31 s by batch sizes 42 and 50. Although gas prices remain moderate—averaging around 5.98 Gwei with a standard deviation of 0.92 Gwei—blocks can reach sizes above 400KB and occasionally include more than 300 transactions, (pushing the network closer to its capacity) \cite{croman2016scaling, buterin2019eip1559}. As with other functions, once blocks approach the 30M gas limit \cite{wood2014ethereum}, validators raise the bar for fee-based prioritization \cite{EthereumProofOfStake}, increasing mempool wait times. Hence, even a comparatively simple Ether-transfer operation can experience notable latency growth under substantial batch concurrency, emphasizing that PoS-based transaction ordering and competition affect all contract calls when block space tightens.

\begin{table*}[t!]
    \centering
    \footnotesize
    \caption{List of Blockchain-Related Terminology Used and Their Definitions}
    \label{tab:blockchain_terms}
    \begin{tabular}{p{4cm} p{12cm}}
        \toprule
        \textbf{Term} & \textbf{Definition} \\
        \midrule
        Arrays & A dynamic or fixed-size list of elements in Solidity, stored either in memory or storage. \\
        Block Size & The amount of data a block can contain, affecting network scalability and congestion. \\
        Cold Writes & First-time storage operations that require additional gas due to state initialization. \\
        Concurrency & The execution of multiple transactions simultaneously, impacting block finalization time. \\
        Dependent Transactions & Transactions requiring a prior transaction to be confirmed before execution. \\
        Enum & A user-defined data type in Solidity that allows defining named constants. \\
        EVM & Ethereum Virtual Machine, the runtime environment for smart contracts on Ethereum. \\
        Finalization Time & The duration required for a block to be confirmed and considered immutable. \\
        Gas Limit & Maximum amount of gas that can be used per block. \\
        Gas Price & Cost per unit of gas in Ethereum, measured in Gwei, fluctuating based on network congestion. \\
        Gas Used & Total gas consumed by an individual transaction, impacting overall transaction costs. \\
        Index & The position of an element in an array, with retrieval costs increasing in deep structures. \\
        Mapping & A key-value store in Solidity, allowing for data retrieval within contracts. \\
        Mempool & Temporary storage for pending transactions before block inclusion. \\
        Nested Array & An array within another array, enabling multi-dimensional storage in Solidity. \\
        Nested Mapping & A mapping that contains another mapping as its value, used for complex data structures. \\
        PoS & Proof of Stake, a consensus mechanism where validators stake assets to validate transactions. \\
        SLOAD & Storage load operation in Ethereum's EVM, retrieving data from contract storage. \\
        SSTORE & Storage write operation in Ethereum's EVM, which incurs higher gas costs than reads. \\
        State Trie & Ethereum's hierarchical data structure that stores account states and contract storage. \\
        Validators & Participants in a PoS blockchain responsible for proposing and validating blocks. \\
        Warm Writes & Subsequent storage operations that consume less gas as they update an existing slot. \\
        \bottomrule
    \end{tabular}
\end{table*}

\begin{table*}[t!]
\centering
\footnotesize
\caption{Summary of Lessons Learned for EVM-Based Smart Contracts in PoS Blockchains for Inter-Provider Agreements: Key Observations, Quantitative Impacts, and Recommendations.}
\label{tab:lessons_learned}
\begin{adjustbox}{max width=1\textwidth}
\begin{threeparttable}
\begin{tabular}{m{2cm} m{5.2cm} m{4.5cm} m{5.5cm}}
\toprule
\textbf{\hspace{5mm} Lessons \hspace{5mm}}
 & \textbf{\hspace{12mm} Key Observations \hspace{12mm}}
 & \textbf{\hspace{12mm} Quantitative Impact \hspace{12mm}}
 & \textbf{\hspace{12mm} Practical Recommendations \hspace{12mm}} \\

\midrule

\multicolumn{4}{c}{\textbf{Smart-Contract Design \& Storage Patterns}} \\ 
\midrule

Avoid Excessive Cold Writes  
 & 
 \begin{itemize}
     \item Cold writes (zero-to-nonzero transitions) require significantly more gas than warm writes.
     \item Large-scale registration events or first-time storage allocations amplify gas overhead.
     \item Concurrent cold writes can saturate blocks, increasing transaction delays.
 \end{itemize}
 & 
 \begin{itemize}
     \item Gas overhead: \(\uparrow\) 17,000 per cold write.
     \item Block usage: \(\uparrow\) 80--90\% capacity under high concurrency.
     \item Latency increase: \(\uparrow\) 10--20s in heavy registration bursts.
 \end{itemize}
 & 
 \begin{itemize}
     \item Merge initial storage steps (e.g., \textit{register + add service}) into a single function call.
     \item Use zero-based enumerations to minimize unnecessary non-zero writes.
     \item Pre-initialize known participants in consortium blockchains to reduce on-demand cold writes.
     \item Schedule large onboarding phases to avoid simultaneous cold-write congestion.
 \end{itemize}
\\
\midrule

Flatten Deep Data Structures 
 & 
 \begin{itemize}
     \item Nested mappings (e.g., \texttt{mapping(address => mapping(...))}) cause repeated \texttt{SLOAD} lookups.
     \item Larger array indices result in higher gas costs due to deeper traversals.
     \item Concurrency amplifies lookup overhead, increasing overall block saturation.
 \end{itemize}
 & 
 \begin{itemize}
     \item Gas cost per deep lookup: \(\uparrow\) 5--10\% per \texttt{SLOAD}.
     \item Multi-level mappings: \(\uparrow\) 10--15\% overhead under concurrency.
     \item Finalization time: \(\uparrow\) 20--30\% when querying nested structures in parallel.
 \end{itemize}
 & 
 \begin{itemize}
     \item Replace nested mappings with single-level hashed keys (e.g., \texttt{bytes32} composed of address + ID).
     \item Partition large arrays into smaller chunks to reduce traversal costs.
     \item Cache repeated lookups in local variables rather than calling \texttt{SLOAD} multiple times.
 \end{itemize}
\\
\midrule

\multicolumn{4}{c}{\textbf{Concurrency \& Transaction Processing}} \\ 
\midrule

Manage Concurrency to Reduce Latency  
 & 
 \begin{itemize}
     \item Larger batch sizes (30--50+) significantly increase transaction finalization times.
     \item Competition for limited block space leads to transaction deferrals.
     \item Mempool prioritization favors higher-gas transactions, delaying others.
 \end{itemize}
 & 
 \begin{itemize}
     \item Increasing batch size from 2 to 50: median finalization time \(\uparrow\) 13s to 30+s.
     \item Extreme outliers at 60--90s due to fee bidding and block contention.
 \end{itemize}
 & 
 \begin{itemize}
     \item Stagger high-load transactions (e.g., randomize registration timing) to prevent burst congestion.
     \item Use priority fees (e.g., +2--3 Gwei) for urgent operations.
     \item Monitor block utilization in real time and queue non-urgent transactions when block occupancy exceeds 80\%.
 \end{itemize}
\\
\midrule

\multicolumn{4}{c}{\textbf{Network-Level Considerations}} \\ 
\midrule

Concurrency-Driven Block Saturation
 & 
 \begin{itemize}
     \item Blocks reach 80--90\% capacity when many gas-intensive transactions occur in a short time.
     \item Some experiments recorded 42 consecutive high-usage blocks.
     \item Partial saturation results in repeated deferrals across multiple blocks.
 \end{itemize}
 &
 \begin{itemize}
     \item Large transaction batches (30--50 operations) increase block congestion.
     \item First-service additions or deeply indexed selections \(\uparrow\) gas usage.
     \item Confirmation time \(\uparrow\) due to repeated near-saturation of blocks.
 \end{itemize}
 &
 \begin{itemize}
     \item Randomize or schedule large registration waves to prevent sustained block pressure.
     \item Break down expensive loops or multi-step operations into smaller sub-transactions.
     \item Coordinate concurrency strategies (from Section~\ref{lesson-2}) with network-level optimizations.
 \end{itemize}
\\
\midrule

Mempool Dynamics \& On-Chain Finalization  
 & 
 \begin{itemize}
     \item Mempool exit times vary: transactions leave quickly but may still experience on-chain delays due to block congestion.
     \item Distinction between \emph{mempool time} (waiting for inclusion) and \emph{on-chain finalization time} (state update delay).
     \item Dependent transactions (e.g., breach $\rightarrow$ penalty) risk multi-block separation.
 \end{itemize}
 &
 \begin{itemize}
     \item Average block delay: \(\uparrow\) 13.38 blocks ($\sim$160s), worst-case 677 blocks in dependent transactions.
     \item 60--90s outliers observed under extreme concurrency peaks.
 \end{itemize}
 &
 \begin{itemize}
     \item Introduce short pauses or separate function calls for sequential transactions to avoid landing in the same congested block.
     \item Monitor block occupancy in real time and defer non-essential writes if utilization remains high.
     \item Shift bulk updates to off-peak hours to reduce both mempool congestion and block saturation.
 \end{itemize}
\\

\bottomrule
\end{tabular}
\begin{tablenotes}
\footnotesize
\item \textbf{Legend:} \(\uparrow\) = Increase in gas cost, latency, or block usage. 
\end{tablenotes}
\end{threeparttable}
\end{adjustbox}
\end{table*}

\section{Lessons Learned and Practical Implications: Insights and Techniques}
\label{lesson-learned}
In this Section, we integrate our empirical findings with design considerations to propose practical guidance. Table \ref{tab:blockchain_terms} presents the terminology used throughout this section and will be referenced throughout this Section.

Table \ref{tab:lessons_learned} summarizes the key lessons learned from implementing smart contracts deployed on EVM using PoS (see Table \ref{tab:blockchain_terms} for definitions) for blockchains for inter-provider agreements. The table presents key observations, quantitative impacts, and practical recommendations related to smart contract design, concurrency (see Table \ref{tab:blockchain_terms}) management, and network-level considerations. These findings are derived from empirical analysis and real-world testing, highlighting the challenges associated with smart contract design and storage patterns (Section \ref{lesson-1}), deep data structures, transaction batching, and network saturation (Sections \ref{lesson-2} and \ref{lesson-3}, respectively).

\subsection{Smart-Contract Design \& Storage Patterns}
\label{lesson-1}
In this subsection, we discuss two critical lessons presented in Table \ref{tab:lessons_learned}, which emerged from our empirical analysis: (i) avoiding excessive cold writes during initial storage allocations and (ii) flattening deep or nested data structures to reduce repeated lookup overhead. These considerations fall under the broader category of smart contract design and data layout optimization. By examining specific quantitative outcomes from our experiments, we highlight why these practices matter in blockchain-based decentralized applications—particularly under multi-domain and concurrency-driven scenarios—and provide practical guidance for designing smart contracts for multi-contract DApps. The key blockchain terms used in this section are defined in Table \ref{tab:blockchain_terms}.

\noindent
\textbf{Avoid Excessive Cold Writes:}
Our results found that cold writes—storing a non-zero value to an uninitialized contract storage slot—consume more gas than warm writes (see Table \ref{tab:blockchain_terms} for its definition), which involve overwriting an already initialized slot. For instance, in registration experiments described in Section \ref{gas-creation-explaination}, initializing a provider’s storage slot from 0 to 1 incurred approximately 17,000 additional gas units. Figure~\ref{fig:gas-registeration} illustrates this phenomenon, showing that the initial provider registration cost around 110,839 gas, compared to 93,739 gas for subsequent registrations. Similar results emerged when adding the first service to a provider’s uninitialized service array. Figure~\ref{fig:gas_addservice} and Table~\ref{tab:gas-addservice} indicate the first service addition consumed roughly 162,500 gas, compared to approximately 146,629 gas for subsequent service additions.

These observations align with technical documentation and prior research. According to the EVM documentation \cite{wood2014ethereum} and studies on storage behavior \cite{ayub2023storage, wackerow2024merkle}, zero-to-nonzero state transitions cause extensive updates to the \textit{state trie} (see Table \ref{tab:blockchain_terms}). In contrast, overwriting an existing non-zero value or resetting a slot to zero may be cheaper due to partial refunds or reduced trie updates.

Concurrency experiments, conducted with batch sizes ranging from 2 to 100 for the function \function{registerAD}, highlighted that cold writes amplify concurrency-related performance issues. Blocks with numerous simultaneous cold writes approach the 30M \textit{gas limit} (see Table \ref{tab:blockchain_terms} for definition) \cite{buterin2014ethereum}, increasing the risk of partial block saturation. For example, concurrent registrations of 44 or more providers raised transaction finalization times (see Table \ref{tab:blockchain_terms} for definition) from approximately 13 s to 25–30 s, influenced by the combined effect of increased gas consumption and network variability.

This cold-write overhead is particularly important in multi-administrative blockchain applications, such as cross-domain resource allocation in future networks (e.g., 6G, beyond-5G) \cite{croman2016scaling}, where multiple administrative domains frequently onboard or update resources simultaneously. Under these conditions, the collective impact of numerous cold writes can saturate blocks (reaching 80–90\% of gas limits), causing transactions to spill over into subsequent blocks and increasing latency, potentially affecting SLAs and on-chain settlements.


\noindent
\textbf{Flatten Deep Data Structures:}
Beyond cold writes, our experiments revealed that data-structure depth and complexity influence gas usage. In scenarios with nested data structures, extra operations to access or modify values increase gas consumption. One scenario involved service selection through nested lookups, where selecting a service with a higher index in an array resulted in more \texttt{SLOAD} (see Table \ref{tab:blockchain_terms}) instructions and increased gas cost \cite{jezek2021ethereum}. Figure~\ref{fig:gas-service-selection} shows that services deeper within the array used 5–10\% more gas, as the EVM traversed additional indices.

A similar pattern occurred with Nested Mapping structures (see Table \ref{tab:blockchain_terms}). Multi-level mappings, for example \texttt{mapping(address => mapping(uint256 => SomeStruct))}, required repeated key lookups, with each nested access incurring an extra \texttt{SLOAD} operation \cite{wood2014ethereum, wackerow2024merkle}. Tests with the \texttt{SelectService} contract indicated that additional lookup layers raised gas costs by 10–15\% during concurrent transactions.

Elevated concurrency increased overhead when multiple consumers queried the same nested structure. Block-level contention pushed blocks toward their gas limits and delayed finalization. Data in Table~\ref{table:selectservice} shows that average block size and transaction count rose when 50 consumer domains selected services concurrently, with finalization times increasing by 20–30\% compared to flat lookups.

Deep or nested data structures raise the cost of routine operations. In a multi-domain environment, each provider may list multiple services, each with sub-resources such as location, cost, or metadata. In heavily nested data—such as a service mapping within another mapping keyed by provider addresses—each read or write becomes a chain of \texttt{SLOAD} (see Table \ref{tab:blockchain_terms}) calls. Under concurrency, these repeated lookups accumulate, and repeated storage reads strain block capacity \cite{ayub2023storage}.

\subsubsection{Practical Suggestions and Techniques}
\label{sec:practical_suggestions}

Below is a unified set of best practices derived from our analyses of cold writes, concurrency management, and blockchain network-level constraints. These measures aim to mitigate block saturation, reduce transaction latency, and enhance the performance of multi-contract DApps (a DApp that is composed of multiple smart contracts, each responsible for distinct functionalities within the system) in public PoS environments.

\begin{enumerate}
    \item Implement real-time analytics to track block occupancy and gas usage \cite{gervais2016security, buterin2019eip1559}. If utilization exceeds a threshold (e.g., 70--80\%) over consecutive blocks, the system can defer or queue non-urgent operations. 
    This approach prevents simultaneous high-gas bursts that push transactions into subsequent blocks.
    \item  Slightly raising gas tips (an extra priority fee (a tip) that users can add to incentivize validators (see Table \ref{tab:blockchain_terms} and Section \ref{sec3} for context on validators in PoS) to process their transaction faster. e.g., 2--3 Gwei) for time-sensitive transactions, such as penalty calculations or urgent SLA breaches, can expedite inclusion when blocks are near full capacity. Developers should limit this strategy to genuinely critical operations and allow routine tasks to proceed with baseline fees.
    \item When registering new domains, adding services, or finalizing penalty transactions, distributing these actions over a multi-minute window helps avoid consistently reaching 80--90\% block usage. Randomizing transaction timestamps or targeting off-peak periods (based on historical or real-time network data) can further minimize contention.
    \item Merging storage-initializing tasks (e.g., registering a provider and adding an initial service) into a single transaction reduces repeated zero-to-nonzero overhead. Although this may raise the gas cost of that one call, it decreases the total number of transactions competing for block space under concurrency.
    \item Replace nested mappings (\texttt{mapping(address => mapping(uint256 => Data))}) with a single-level key (\texttt{mapping(bytes32 => Data)}) to reduce repeated lookups. Similarly, partition large arrays into smaller segments or sub-arrays to avoid excessive indexing costs during concurrent access \cite{ayub2023storage}.
    \item For expensive loops or deeply nested updates, split the operation into smaller batches. For instance, adding 20 services in groups of five helps keep each transaction below critical gas thresholds, diminishing the risk of block saturation.
    \item In environments with frequent on-boarding, assigning the most common role (e.g., “Consumer=0”) as a default Enum (see Table \ref{tab:blockchain_terms}) minimizes unnecessary non-zero writes. For known participants, pre-allocating storage slots can also smooth out spikes from simultaneous registrations \cite{buterin2019eip1559}.
    \item Repeated \texttt{SLOAD} operations can be costly, especially under concurrency. Caching data in local variables (e.g., reading a mapping entry once and updating it in memory) reduces repeated lookups \cite{jabbar2020investigating}.
\end{enumerate}

By combining these strategies, development of smart contract can mitigate partial block saturation, reduce gas overhead from cold writes, and maintain more predictable transaction latencies. Although no single approach fully eliminates concurrency-induced delays, applying a combination of these techniques can significantly improve performance for multi-contract DApps operating in public PoS networks.



\begin{table}[t!]
    \centering
    \caption{Block Inclusion Metrics for Varying Batch Sizes in High Gas Usage Scenario (\function{addService})}
    \label{tab:batch_size_metrics}
    \begin{tabular}{l c c}
        \toprule
        \textbf{Batch Size} & \textbf{Distinct Blocks} & \textbf{Avg. Transactions per Block} \\
        \midrule
         2  & 10 & 127.90 \\
        10  & 10 & 115.30 \\
        18  & 17 & 129.12 \\
        26  & 19 & 136.47 \\
        34  & 24 & 126.42 \\
        42  & 26 & 134.65 \\
        50  & 29 & 121.72 \\ 
        \bottomrule
    \end{tabular}
    \vspace{0.5em}
    
    \begin{minipage}{0.9\textwidth}
      \small \textit{Note:} The “Distinct Blocks” column indicates the number of unique blocks in which the transactions were included for each batch size. The “Avg. Transactions per Block” provides an indication of the block fullness during these operations.
    \end{minipage}
\end{table}

\begin{table}[t!]
\centering
\small
\caption{Gas Usage and Block Congestion Insights. ``Cumulative Gas Used (B)'' is reported in billions of gas units, while ``DApp Gas (M)'' is in millions. Each block is capped at $\sim$30M gas.}
\label{tab:gas_usage_block-saturation}
\begin{tabular}{l c c c c}
    \toprule
    \textbf{DApp functions} & 
    \textbf{Cumulative Gas (B)} & 
    \textbf{DApp Gas (M)} & 
    \textbf{Percentage of Chain} & 
    \textbf{Blocks Below 80\%} \\
    \midrule
    \function{registerAD} & 62.05 & 133.89 & 0.22\% & 42 \\
    \function{addService}   & 28.88 & 133.14 & 0.46\% & 13 \\
    \function{serviceSelection}& 19.85 & 152.59 & 0.77\% & 13 \\
    \bottomrule
\end{tabular}
\vspace{5pt}
\begin{minipage}{0.9\textwidth}
\footnotesize
\textit{Note:} \textbf{Cumulative Gas (B)} represents the total gas consumed by each transaction type across all observed blocks in billions of gas units. \textbf{DApp Gas (M)} refers to the gas consumed per transaction type in millions of gas units. \textbf{Percentage of Chain} indicates the proportion of total on-chain gas usage attributed to each operation. Blocks Below 80\% denotes the number of blocks where gas usage remained under 80\% of the 30M gas limit. No block exceeded the 30M limit in these experiments.
\end{minipage}
\end{table}

\subsection{Concurrency \& Transaction Processing Time}
\label{lesson-2}
This Section discusses how multiple administrative domains acting in parallel can affect transaction finalization times on a PoS blockchain. Table \ref{tab:lessons_learned} presents a summary of key observations and provides practical recommendations to address the challenges highlighted in our analysis. The analysis shows that competition for finite block space remains a bottleneck, even when the network operates with more predictable block intervals than under a PoW consensus model. Based on our evaluations (e.g., large batch sizes inducing concurrency-driven delays), we outline key findings and practical recommendations to mitigate latency spikes in high-throughput inter-provider scenarios.

\noindent
\textbf{Manage Concurrency to Reduce Latency:}
Our experimental results show a correlation between batch size (the number of administrative domains acting in parallel) and transaction latency. When registration or service addition occurs in batches of 2–4 participants, median finalization times remain below 15 s, matching the expected block interval in Ethereum PoS \cite{EthereumProofOfStake}. As batch sizes increase to 30, 50, or 100, average latency rises above 30 s, with some cases exceeding 60–90 s.

These delays result from block-level contention. Ethereum’s PoS mechanism enforces a gas limit of about 30 million per block \cite{wood2014ethereum, buterin2014ethereum,javed2025empiricalsmartcontractslatency}. When many domains issue storage-heavy operations—such as adding an uninitialized service, registering a cold-write provider role, or updating nested data—blocks fill faster, pushing transactions into later blocks. Figures in Section \ref{sec:concurrency-latency} (e.g., Figure \ref{fig:Latency-box-registeration} and Table \ref{table-registeration-full-details}) show that as concurrency increases, transaction counts per block approach the gas capacity, and finalization latencies rise.

Empirical data also shows that mempool dynamics under concurrency do not penalize large transactions uniformly. When transactions with similar gas tips compete, higher-gas calls are included as block space becomes available. They exit the mempool quickly but may be deferred on-chain if the block is partially saturated \cite{gencer2018decentralization}. This leads to moderate mempool wait times alongside delayed confirmations across multiple blocks. Actions such as registering multiple breaches (Section \ref{fig:boxplot-write-breach-max}) or selecting services in unison (Section \ref{fig:boxplot-write-selection}) show increased latencies. Overall, concurrency pushes transactions into later blocks and increases processing times \cite{javed2025empiricalsmartcontractslatency}.


Network resource competition affects system performance beyond individual transactions. The block’s capacity limits the number of concurrent transactions processed in each interval, even when gas bids are competitive. This limit leads participants to raise gas tips to secure inclusion, which increases overall fees \cite{buterin2019eip1559}. High concurrency can saturate network blocks, affecting other contracts on the same network, and excessive activity in one application may impact the blockchain when aggregated gas usage is significant.

Concurrency challenges extend beyond local performance issues and may disrupt time-sensitive workflows while intensifying competition in the fee market. In multi-provider 6G use cases that depend on on-demand interactions, these issues can undermine dynamic resource sharing and call for strategies to reduce latency spikes.

\paragraph{Practical Suggestions and Techniques}
\begin{enumerate}
    \item Instead of broadcasting large batches (e.g., 50 or 100 providers) simultaneously, schedule or randomize submission times over a short window (e.g., one minute). This prevents bursty traffic that quickly saturates consecutive blocks and defers slower or lower-tipped transactions.
Refer to Section~\ref{lesson-1} for techniques on combining cold writes into fewer transactions, which also reduces concurrency pressure.
    \item Raising gas tips (e.g., by 2–3 Gwei) can help critical transactions—like SLA breach penalties—gain priority inclusion under heavy load. However, routine actions should keep standard fees to avoid network-wide inflation \cite{buterin2019eip1559}. Earlier sections discuss additional methods for flattening data structures and caching repeated lookups, which further lower the gas load under concurrency.
\end{enumerate}

By adopting these concurrency management strategies, developers can mitigate the sharp latency spikes observed in batch-driven load. The interplay between PoS validators, finite block gas capacity, and partially uniform gas tips means that concurrency remains a critical bottleneck: even modest concurrency growth can lead to disproportionately longer confirmation delays \cite{wood2014ethereum, gencer2018decentralization}. Hence, a proactive approach—staggering large updates, dynamically adjusting fees, and focusing on off-peak intervals—facilitates smoother system performance in PoS live environments. 

\begin{table}[t!]
    \centering
    \caption{Block Delay Metrics for Dependent Transaction Processing}
    \label{tab:delay_stats_updated}
    \begin{tabular}{l c}
        \toprule
        \textbf{Metric} & \textbf{Value} \\
        \midrule
        Total Delayed Cases & 712 \\
        Average Block Delay (blocks) & 13.38 \\
        Median Block Delay (blocks) & 5.00 \\
        Maximum Block Delay (blocks) & 677 \\
        90th Percentile Block Delay (blocks) & 14.00 \\
        \midrule
        \multicolumn{2}{l}{\small \textit{Note: At 12-second block intervals, the average delay corresponds to approximately 160 seconds.}} \\
        \bottomrule
    \end{tabular}
\end{table}

\subsection{Network-Level Considerations}
\label{lesson-3}
This subsection explores how transaction concurrency affects block saturation and mempool behavior in public PoS blockchains. Despite more predictable block intervals than PoW, limited block capacity and fee-based prioritization can lead to multi-block delays during heavy usage. By referencing quantitative data and theoretical insights, we highlight the core issues—block saturation, mempool constraints—and practical steps to address them. Table~\ref{tab:lessons_learned} also provides a summary of key observations and actionable suggestions.

\noindent
\textbf{Concurrency-Driven Block Saturation:}
Our empirical observations, drawn from multiple experiments, illustrate how block saturation can occur when numerous gas-intensive operations converge in a short timeframe. While each transaction remains below the per-block gas limit (\(\sim\)30\,M gas on Ethereum), their cumulative effect can push consecutive blocks near capacity. Table~\ref{tab:gas_usage_block-saturation} highlights one representative dataset showing how functions (\function{registerAD}, \function{addService}, \function{serviceSelection}) contributed to chain usage. Notably, in this instance, \function{serviceSelection} logs the lowest total gas (19.85B) yet the highest ``Percentage of Chain'' (0.77\%), because it often ran during lower global usage. In contrast, \function{registerAD} consumed more total gas (62.05B) but only reached 0.22\%, as many of its transactions landed in blocks already congested by other activities.

Another useful metric in Table~\ref{tab:gas_usage_block-saturation} is “Blocks Below 80\%,” indicating how often blocks stayed under 80\% of the 30\,M limit for each function. Fewer blocks below 80\% suggests that particular operations tended to push block usage closer to saturation—consistent with concurrency surges. Although none of these functions alone reached the 30\,M cap, partial saturation still forced some transactions into subsequent blocks. Even smaller-scale calls can significantly affect congestion if they cluster in time or land during otherwise quiet periods.

Moreover, \emph{cold writes} (transitioning storage from zero to non-zero) add extra overhead \cite{wood2014ethereum, ayub2023storage}. Under concurrency, these frequent cold writes can push utilization to 80--90\% of the limit. Our measurements recorded 42 consecutive blocks near capacity due to repeated \function{registerAD} bursts. While no single block hit 100\%, the sustained pressure caused a phenomenon of \emph{repeated partial saturation}, where transactions spilled over into subsequent blocks and extended confirmation times. Figures~\ref{fig:box-1addService-2-50x10}, \ref{fig:CDF-1addService-2-50x10}, and Table~\ref{tab:batch_size_metrics} further show how batch sizes from 2 to 50 required more blocks to handle the load, lengthening end-to-end latency. Similarly, the first time a service is added (a cold write) or a deeply indexed service is selected, the higher gas cost under concurrency makes it easier to approach block capacity.

\noindent
\textbf{Mempool Dynamics \& On-Chain Finalization:}
Under uniform gas tips, larger or more complex transactions do not necessarily linger in the mempool; they are included once sufficient block space becomes available \cite{gencer2018decentralization}. If a block is near capacity, the inclusion of any pending transaction—regardless of size—may be deferred to the next block. Consequently, there is a distinction between \emph{mempool time} (waiting for a validator to pick it) and \emph{on-chain finalization time} (potentially spanning multiple near-full blocks). As concurrency rises, repeated partial saturation inflates average confirmation times from about 13--15\,s to 30\,s or more, with occasional outliers at 90\,s \cite{croman2016scaling, buterin2019eip1559}. Table~\ref{tab:delay_stats_updated} further quantifies these multi-block deferrals in dependent transactions (e.g., registering a breach before calculating penalties), showing an \emph{average delay of 13.38 blocks} ($\sim$160\,s at 12\,s/block) and a maximum of 677 blocks. Such outliers typically arise during large batch experiments or when numerous cold writes and nested lookups coincide, saturating several blocks in succession. In these scenarios, \emph{block occupancy}—rather than mempool ordering—ultimately dictates when all state updates finalize on-chain, mirroring other blockchain performance research \cite{croman2016scaling, buterin2019eip1559}.


These findings are crucial for multi-domain DApp deployments, especially in 6G or beyond-5G environments requiring near-real-time resource sharing. Several challenges emerge:

\begin{itemize}
    \item Even if mempool wait times are modest, partial saturation can postpone transactions by multiple blocks. With an average block interval of $\sim$12 seconds \cite{EthereumProofOfStake}, this adds tens of seconds to final confirmation—acceptable in some contexts, but prohibitive when delays approach 60--90 seconds \cite{croman2016scaling}.
    \item Under higher concurrency, participants raise gas tips to ensure priority, potentially sidelining smaller or cost-sensitive stakeholders \cite{buterin2019eip1559}.
    \item A single DApp’s bursts can affect unrelated contracts by filling blocks and raising baseline gas prices (see Table \ref{tab:blockchain_terms} for its definition). External events can similarly reduce available block space.
    \item Dependent calls (e.g., breach registration followed by penalty calculation) can land in different blocks if the first call saturates the block. In extreme peaks, we observed multi-block separations exceeding 1--2 minutes. As summarized in Table~\ref{tab:delay_stats_updated} presents delay metrics measured in blocks, with one block corresponding to a 12-second interval. On average, dependent transactions are delayed by 13.38 blocks—roughly 160 seconds. In more extreme cases, delays reached up to 677 blocks, highlighting the potential for multi-block deferrals under heavy concurrency.
    \item Incremental increases in concurrency can cause sizable jumps in block usage once it surpasses 70--80\%. Large batch sizes (50--100 transactions) often saturate several consecutive blocks, lengthening finalization for all included transactions.
\end{itemize}


\paragraph{Practical Suggestions and Techniques}
Addressing block saturation and mempool dynamics requires a combination of contract design, operational scheduling, and careful fee strategies. While no single method eliminates all concurrency-induced delays, the following recommendations can mitigate their impact:

\begin{enumerate}
    \item Continuously monitor block occupancy and mempool size. If usage exceeds a threshold (e.g., 70–80\%) for multiple intervals, automatically queue or delay non-critical operations to prevent sustained partial saturation \cite{gervais2016security, buterin2019eip1559}. Refer to Section~\ref{lesson-2} for suggestions on scheduling large batches and using targeted fee increases, which dovetail with network-level throttling.
    \item Encourage a short pause or second function call between related operations (e.g., breach registration $\rightarrow$ penalty calculation) to avoid both steps landing in the same congested block. This approach reduces multi-block separation for sequential tasks, particularly under extreme concurrency peaks. Earlier concurrency guidance (Section~\ref{lesson-2}) also applies if multiple domains attempt the same multi-step workflow in parallel.
\end{enumerate}

Empirical data consistently indicates that repeated partial saturation—rather than isolated large transactions—drives multi-block deferrals, raising end-to-end confirmation times even when mempool exit occurs swiftly. In next-generation resource-sharing scenarios, where multiple administrative domains frequently register, select services, or log SLA breaches in parallel, concurrency can boost throughput but also extend latencies under PoS block intervals. Subtle differences in gas tips yield minor advantages, yet fundamental capacity limits persist whenever several gas-intensive calls collide in the same block.

Network-level considerations thus require deliberate scheduling and, in some cases, rethinking contract calls. Staggering high-registration waves, segmenting expensive subroutines, adjusting fee strategies, and tracking block usage can improve resilience against partial saturations. However, in a decentralized setting lacking a central scheduling authority, these measures depend on coordination or incentives. Although observed saturation rarely surpasses the 30-million-gas ceiling in one shot, repeated events commonly force transactions into subsequent blocks, producing multi-block latency expansions. Concurrency-aware design is therefore crucial for multi-domain or cross-domain frameworks, allowing participants to better control finalization times, alleviate network congestion, and maintain stable performance.

\subsection{Summary of Lesson Learned: Key Takeaways}

Taken together, these findings indicate that effective on-chain performance is not solely a matter of ``efficient'' code implementation. It also relies on coordinating transaction timing, optimizing data layouts, and managing concurrency in public blockchain environments. Each phase of the experiments demonstrates that cost and latency are shaped by several intertwined factors, including the overhead of cold writes, the complexity of nested data structures, and the capacity constraints introduced by PoS consensus. Gas costs consistently rise whenever the system performs non-zero initialization in previously uninitialized slots, and latencies become more pronounced when multiple administrative domains submit transactions simultaneously.

The results show that even though a single registration event might have minimal effects, large bursts of newcomers—each with its own non-zero state transitions—push blocks toward their gas ceiling. This phenomenon is evident when many providers join at once, causing block saturation and eventually deferring some transactions to subsequent blocks. Similarly, deeply nested arrays  or mappings (see Table \ref{tab:blockchain_terms}) incur overhead for repeated \texttt{SLOAD} operations, a cost that compounds when multiple users read or write to the same or related structures. Flattening these data layouts through less nesting—or caching intermediate results in local memory—can mitigate per-transaction overhead.

Concurrency amplifies these trends by increasing the rate of incoming transactions. When larger batches of administrative domains act concurrently, partial block saturation arises across multiple blocks, while higher gas bids yield only incremental gains in transaction priority. These bottlenecks extend finalization beyond typical block intervals, hampering time-sensitive procedures such as service-level agreement enforcement. Scheduling transactions in off-peak times, introducing dynamic fee adjustments, or bundling certain actions into single function calls can mitigate the effects of these concurrency waves.

From a network-level perspective, mempool selection alone is insufficient to control transaction throughput. Once the blockchain nears capacity, even transactions that exit the mempool quickly might be deferred to later blocks, reflecting the limited block space in each interval. This underscores the need for developers and system architects to anticipate how concurrency loads and repeated writes combine to saturate block utilization.

These insights imply that multi-domain architectures, particularly those aimed at future 6G or resource-sharing environments, should incorporate strategies for reducing cold writes, flattening data structures, and smoothing out bursts of high-demand activity. Taken together, these measures ensure that spikes in operations do not unnecessarily raise costs or create excessive delays. Overall, the practices presented—combining initial writes, using shallow mappings, scheduling large transactions during lower-demand periods, and offering priority fees for critical calls—provide a balanced method to maintain performance and predictability under typical network conditions. By adopting these techniques, inter-provider DApps can address real-world performance concerns, confirm transactions more reliably, and minimize the on-chain expenses that often accompany concurrency.

\section{Conclusions}
\label{sec6}
In this work, we designed, implemented, and empirically evaluated a multi-contract DApp for inter-provider agreements on the Ethereum Sepolia testnet, providing detailed insights into gas consumption, transaction latency, and block-level resource utilization under realistic network conditions. Our experimentation revealed that repeated cold writes introduce a 15–20\% increase in gas usage—crucially affecting large onboarding phases when new domains register en masse. Likewise, we observed that deep or nested data structures can impose a 5–10\% overhead in gas costs per access, magnified when multiple participants act concurrently. As batch sizes rose, finalization times grew from baseline levels near 13–15 seconds to over 30 seconds or more, especially when block usage approached 80–90\% of its gas limit.

These findings affirm that gas and latency behavior hinge not just on per-function optimizations but also on concurrency management and network-level constraints. Even moderate concurrency waves can saturate blocks, defer some transactions to subsequent blocks, and inflate end-to-end settlement times. Consequently, our lessons learned emphasize (i) avoiding excessive cold writes by consolidating or batching initial state updates, (ii) flattening nested data layouts to reduce repeated lookups, and (iii) scheduling or throttling large transaction bursts so they do not overlap and worsen block congestion. Introducing dynamic fee adjustments for critical operations can further mitigate latencies under load.

Overall, this study demonstrates how a careful interplay of smart-contract design, data-structure choices, and transaction timing can meaningfully improve performance in a multi-administrative 6G environment. By adopting the recommended practices—such as caching, zero-based enumerations, and off-peak scheduling—smart contract development can reduce gas overhead and latency, thereby enhancing the viability of on-chain resource sharing, SLA enforcement, and cross-domain coordination in next-generation telecom systems.

\section*{Acknowledgment}
This work partially funded by Spanish MINECO grants TSI-063000-2021-54/-55 (6G-DAWN), Grant PID2021-126431OB-I00 funded by MCIN/AEI/ 10.13039/501100011033 and by “ERDF A way of making Europe” (ANEMONE), and Generalitat de Catalunya grant 2021 SGR 00770.


\begin{thebibliography}{10}
\bibitem{transactiontimes}
M.~Alharby, A.~Moorsel, Is my transaction done yet? an empirical study of transaction processing times in the ethereum blockchain platform, IEEE Transactions on Blockchain.
\end{thebibliography}
\end{document}